\documentclass[12pt]{article}
\usepackage{latexsym}
\usepackage{epsfig}

\usepackage{graphicx}
\usepackage{epstopdf}

\usepackage{epsfig,amssymb,amsmath,amscd}

\newcommand{\bq}{\begin{equation}}
\newcommand{\eq}{\end{equation}}
\newcommand{\bqa}{\begin{eqnarray}}
\newcommand{\eqa}{\end{eqnarray}}
\newcommand{\ben}{\begin{enumerate}}
\newcommand{\een}{\end{enumerate}}
\newcommand{\bc}{\begin{center}}
\newcommand{\ec}{\end{center}}
\newcommand{\bqb}{\begin{eqnarray*}}
\newcommand{\eqb}{\end{eqnarray*}}

\def\pr#1#2#3{Phys. Rev. ${\bf{#1}}$, #2 (#3)}

\def\pl#1#2#3{Phys. Lett. ${\bf{#1}}$, #2 (#3)}

\def\np#1#2#3{Nucl. Phys. ${\bf{#1}}$, #2 (#3)}

\def\epj#1#2#3{Eur. Phys. J. ${\bf{#1}}$, #2 (#3)}

\def\jmp#1#2#3{J. Mod. Phys. ${\bf{#1}}$, #2 (#3)}

\begin{document}
\pagenumbering{arabic}
\thispagestyle{empty}
\def\thefootnote{\fnsymbol{footnote}}
\setcounter{footnote}{1}

\begin{flushright}
August 14, 2017\\
 \end{flushright}

\begin{center}
{\Large {\bf Tests of bottom quark CSM properties}}.\\
 \vspace{1cm}
{\large F.M. Renard}\\
\vspace{0.2cm}
Laboratoire Univers et Particules de Montpellier,
UMR 5299\\
Universit\'{e} Montpellier II, Place Eug\`{e}ne Bataillon CC072\\
 F-34095 Montpellier Cedex 5, France.\\
\end{center}

\vspace*{1.cm}
\begin{center}
{\bf Abstract}
\end{center}

We look for possible (partial) compositeness properties of
the bottom quark according to the Composite Standard Model (CSM)
concept in the same way this was previously done for Higgs boson and top quark.
After looking at the direct form factor effects appearing in $e^+e^-\to b\bar b$
we analyze the more complex features of the  
$e^+e^-, gg, \gamma\gamma \to b\bar b H, b\bar b Z, t\bar b W^-$ processes.
We emphasize typical differences appearing between CSM concerving and CSM violating cases. 
We also mention the possible appearence of an effective bottom mass.

\vspace{0.5cm}
PACS numbers:  12.15.-y, 12.60.-i, 14.80.-j;   Composite models\\

\def\thefootnote{\arabic{footnote}}
\setcounter{footnote}{0}
\clearpage

\section{INTRODUCTION}

We have recently analyzed the possible tests of the concept of
Compositeness Standard Model (CSM), see ref.\cite{CSMrev},
using several production processes of Higgs boson and top quark.\\

Compositeness of the top quark and of the Higgs boson 
has been studied in \cite{partialcomp,Hcomp2,Hcomp3,Hcomp4}.\\

But our concept consists in assuming that the SM can be constructed, modified
or completed in a way which preserves its main structures and properties
at low energies. One possibility could be its generation
from substructures; for examples of substructures see ref.\cite{comp}.\\

In our analyses we assume that there exist no anomalous coupling which
would already produce a deviation from SM predictions at low energy.
We caracterize our tests by the presence of form factors which affect
the basic SM couplings progressively with the energy but keep the global
structure of the amplitudes. This is particularly important for processes
with production of longitudinal gauge bosons due to the necessary cancellations
of partial elements which would otherwise increase with the energy and
violate unitarity. The preservation of the Goldstone equivalence is a CSM property
which would automatically ensure these cancellations.\\

We have illustrated the differences between CSM conserving and CSM violating
form factor effects for various processes involving the
Higgs boson and/or the top quark; see references in \cite{CSMrev}.
They are often very spectacular.\\

We now consider the possibility of bottom quark compositeness.
The much smaller mass than the one of top quark raises the question
of partial or full bottom compositeness.
As discussed for example in ref.\cite{Hcomp4} there is the
possibility of mixing of elementary states with composite ones.
We will illustrate the two extreme cases with zero and full mixing.\\
Another possibility considered in the top quark case is the formation
of an effective (scale dependent) top mass especially in the case of
pure $t_R$ compositeness. We will apply the same consideration for
the bottom case.\\
The presence of bottom form factors could obviously be first detected in
the simple $e^+e^- \to b\bar b$ process, but we will show how the detailed
CSM properties could be studied in the more involved processes
$e^+e^-, gg, \gamma\gamma \to b\bar b H, b\bar b Z, t\bar b W^-$.\\

Contents: In Section 2 we recall the features of the CSM description.
The analyses of the various above processes are made in Sect.3.
A final summary is given in Sect.4.\\

\section{CSM description}

In our previous papers refered in \cite{CSMrev}, we have established an effective 
description of substructure effects with what we call the CSM concept. 
It consists in assuming that some compositeness model may exist in which
the pure SM is preserved at low energy with its usual set of basic
couplings. We have yet no precise such model allowing a direct computation
of observable effects. But with this concept
no anomalous coupling creating immediate deviation from SM should appear.
The spatial extension due to compositeness
would only generate an energy dependence of the point-like couplings
which means a form factor affecting them, but being
close to 1 at low energy, and controlled at high energy by a
new physics scale related to the binding of the constituents.\\
An example of "test form factor" that we use
in our illustrations is:
\bq
F(s)={s_0+M^2\over s+M^2}~~\label{FF}
\eq
\noindent
with the new physics scale $M$ taken for example in the few TeV range.\\

Such form factors had been affected to the Higgs and to the top quark
with various types of relations; those which satisfy the CSM constraints
(CSMFF) and those which violate them (CSMvFF). In addition one may assume that
Goldstone equivalence is preserved in some effectice manner by CSM and we will
denote these cases as CSMGFF; in practice this means that the amplitudes for longitudinal
gauge bosons can be replaced by amplitudes for Goldstone bosons with the same form factors.
The notations "FF" refers to symbols representing the sector affected by compositeness.
We now list the choices previously made for the top case that we 
extend to the bottom case.\\
In processes involving both top and bottom like in $t\bar b W^-$ production
we will, for simplicity, make the same choices for both of them.\\

CSMbLR and CSMGbLR: $F_{b_R}(s)=F_{b_L}(s)=F(s)$
 and $F_G(s)=F_H(s)=F(s)$
 keeping the bottom mass at its bare value,\\
 
CSMbR and CSMGbR: $F_{b_L}(s)=1$ $F_{b_R}(s)=F(s)$
 and $F_G(s)=F_H(s)=1$, with the effective bottom mass
 $m_b(s)=m_bF(s)$,\\
 
CSMvb: different form factors for $b_L$ (ex: M= 10 TeV)
 and for $b_R$ (ex: M= 15 TeV), and $F_G(s)=F_H(s)=F(s)$,
 keeping a bare bottom mass,\\
 
CSMvH: no bottom form factor but $F_G(s)=F_H(s)=F(s)$
 and the bare bottom mass.\\

\underline{Bottom mixing}\\

The above list refers to the case of full bottom compositeness.
But one should worry about the possibility of partial bottom quark compositeness.
In practice it means for example that the effective $\gamma bb$ and $Zbb$ couplings
should be modified by a factor of the type
\bq
\cos\phi+F(s)\sin\phi
\eq
where $\phi$ is the mixing angle of the elementary bottom quark with
the new sector (which is equal to $\pi/2$ or 0 in the case of full or no
compositeness) and $F(s)$ a form factor similar to the one we used above.\\

In the illustrations we will consider the two extreme cases with zero or with full mixing.\\

\section{Studied Processes}

\underline{$e^+e^-\to b\bar b $}\\

The trivial process for a direct detection of $\gamma b\bar b$
and $Z b\bar b$ form factors is $e^+e^- \to b\bar b$.\\
Using the standard expressions of the polarized or unpolarized
cross sections and asymmetries (see for ex. \cite{eebb}) one can see
the effects of left and right form factors affecting
the bottom couplings.\\ 
In Fig.1 we illustrate the differences
between the effects of pure left ($bL$) , pure right ($bR$) (as in CSMbR), equal left and
right ($bLR$) (as in CSMbLR), and different left and right ($LDR$)
(as in CSMvb) form factors.\\ 
We consider their effects in the unpolarized cross section ($\sigma_{unp}$),
its forward-backward asymmetry ($A^{FB}_{unp}$), the longitudinally polarized
cross sezction  ($\sigma_{long}$), its polarization asymmetry ($A^{FB}_{long}$), 	
also its forward-backward asymmetry ($A^{FB}_{long}$)
and the transverse polarization coefficient  ($\sigma_{trans}$); 
see \cite{trcomp} for the same study in the case of  $e^+e^- \to t\bar t$.\\
It seems clear from Fig.1 that such measurements could allow to determine
which type of form factors could be present.\\

Assuming that the presence of a form factor is detected, the next step
will be to check if it satisfies the CSM requirements.\\
This requires more detailed tests of the complete structure of a possible 
Higgs, top and bottom compositeness.
We will now analyze the informations that could be obtained 
from the study of the
$e^+e^-, gg, \gamma\gamma \to b\bar b H, b\bar b Z, t\bar b W^-$
processes.\\

The involved SM diagrams are recalled in Fig.2-4 and the effects 
of form factors defined in Section 2 are illustrated in Fig.5-13.\\

\underline{$e^+e^-, gg, \gamma\gamma \to b\bar b H$; Fig.5-7}\\

With zero mixing (elementary $b$) one only gets a decreasing effect due to the
presence of the Higgs compositeness form factor appearing in the choices
CSMvb, CSMvH, and CSMbLR.\\

In the full $b$ compositeness case one observes the superposition of $b_L$
and/or $b_R$ and Higgs compositeness decreasing effects.\\

\underline{$e^+e^-, gg, \gamma\gamma \to b\bar b Z$; Fig.8-10}\\

For these processes we make separate illustrations for pure $Z_L$ 
production and for unpolarized $Z$ production.\\
The $Z_L$ cases lead always to clearer results than the unpolarized cases 
but give more or less the same informations.
With zero mixing (elementary $b$) there is essentially no visible effect
except a small one in the pure $Z_L$ case due to the form factor of the Goldstone coupling;
so we do not show the corresponding ratios.\\
With full b compositeness the addition of Higgs anf bottom (left/right) form
factors lead to specific different decreases with the energy as one can see
for CSMvH, CSMvb, CSMbR and CSMbLR choices.\\
Specific quantitative effects can be observed in Fig.8-10 for each of the three 
($e^+e^-, gg, \gamma\gamma$) initial states.\\

\underline{$e^+e^-, gg, \gamma\gamma\to  t\bar b W^-$; Fig.11-13}\\

These processes lead to richer sets of informations because of the possible simultaneous
presence of Higgs, top and bottom compositeness effects possibly consistent with CSM.\\
For simplicity we will make the same assumptions for the choices of top and bottom form
factors but with the 2 possibilities of zero or full $b$ mixing. The CSMG assumption
of validity of the Goldstone equivalence for $W^-_L$ will now be added in the
$bLR$ and $bR$ choices.\\
Separate illustrations are given for $W^-_L$ and for unpolarized $W^-$ production.\\
As for the $Z_L$ case the $W^-_L$ production ratios give clearer but rather similar
informations than the unpolarized ones.\\
We can notice the specific behaviours of the CSM Goldstone choices.\\
Globally the bottom compositeness effects are important and modify
the shapes predicted in the pure top and Higgs compositeness cases, especially in the
gluon-gluon and photon-photon processes and somewhat less in $e^+e^-$ because
the important $W$ emission from $e^{\pm}$ lines is not affected by the
form factors.\\

\underline{An effective bottom mass?}\\

The possibility of an effective top mass generated by compositeness
was mentioned in \cite{trcomp}. This would preserve the CSM concept.
Large effects can be generated that way in all processes where the
top mass plays an important role (especially in the ones involving
Higgs and/or longitudinal gauge bosons).\\

Our question is now if and where the bottom mass could play a similar role
and reveal the presence of an effective scale dependence.\\

A direct effect can be observed in $b\bar b H$ production due to the 
coupling proportional to $m_b$, as we have already seen in Fig.6.\\

Similar effects should be present in $b\bar b Z_L$
(from the Goldstone equivalence $G^0=Z_L$ with the coupling  proportional to $m_b$)
but the  $Z_L$  rate in the total $Z$ production becomes very small at high energy
and will be unobservable if $m_b(s)$ leads to an additional decrease.
Other processes like $Z_LZ_L\to b\bar b$ (equivalent to $G^0G^0\to b\bar b$)
should also be sensitive to $m_b$, however they would be difficult to identify.\\

The effect of $m_b(s)$ in the $t\bar b W^-$ production processes should also be 
negligible because the main mass dependent terms are those of the top quark
which largely hide the bottom ones.\\

So we do not yet see other means of checking the existence of an effective  $m_b(s)$
bottom mass at high energy.\\

\section{Summary}

In this paper we have applied the concept of Composite Standard Model (CSM)
to the bottom quark. The point was to look for tests of (partial or full)
$b$ compositeness and to check if it can be consistent with CSM
in the same way this was done for possible top and Higgs compositeness.\\

We use the test form factors with typical choices of CSM conserving 
and CSM violating combinations similar to those of the top case.\\

We have first looked at the $e^+e^-\to b\bar b $ process in order to
show the modifications of the cross sections and asymmetries that such
form factors would generate.\\

We illustrate then how the 9 processes 
$e^+e^-, gg, \gamma\gamma \to b\bar b H, b\bar b Z, t\bar b W^-$
could allow to study the CSM properties of these form factors.\\

Indeed large and specific differences between the various CSM conserving 
and CSM violating choices appear in the illustrations.\\
The possibility of observing such effects at future high endrgy colliders 
should then be studied; for recent reviews, see for example
\cite{Moortgat,Denterria,Craig} for $e^+e^-$, \cite{Contino} for proton-proton
and  \cite{gammagamma} for photon-photon.\\

Finally we discussed the possibility of the presence of an effective scale 
dependent bottom mass but it seems difficult to observe it because of the small 
effects it would produce in the considered processes.\\

\newpage

\begin{figure}[p]
\[
\epsfig{file=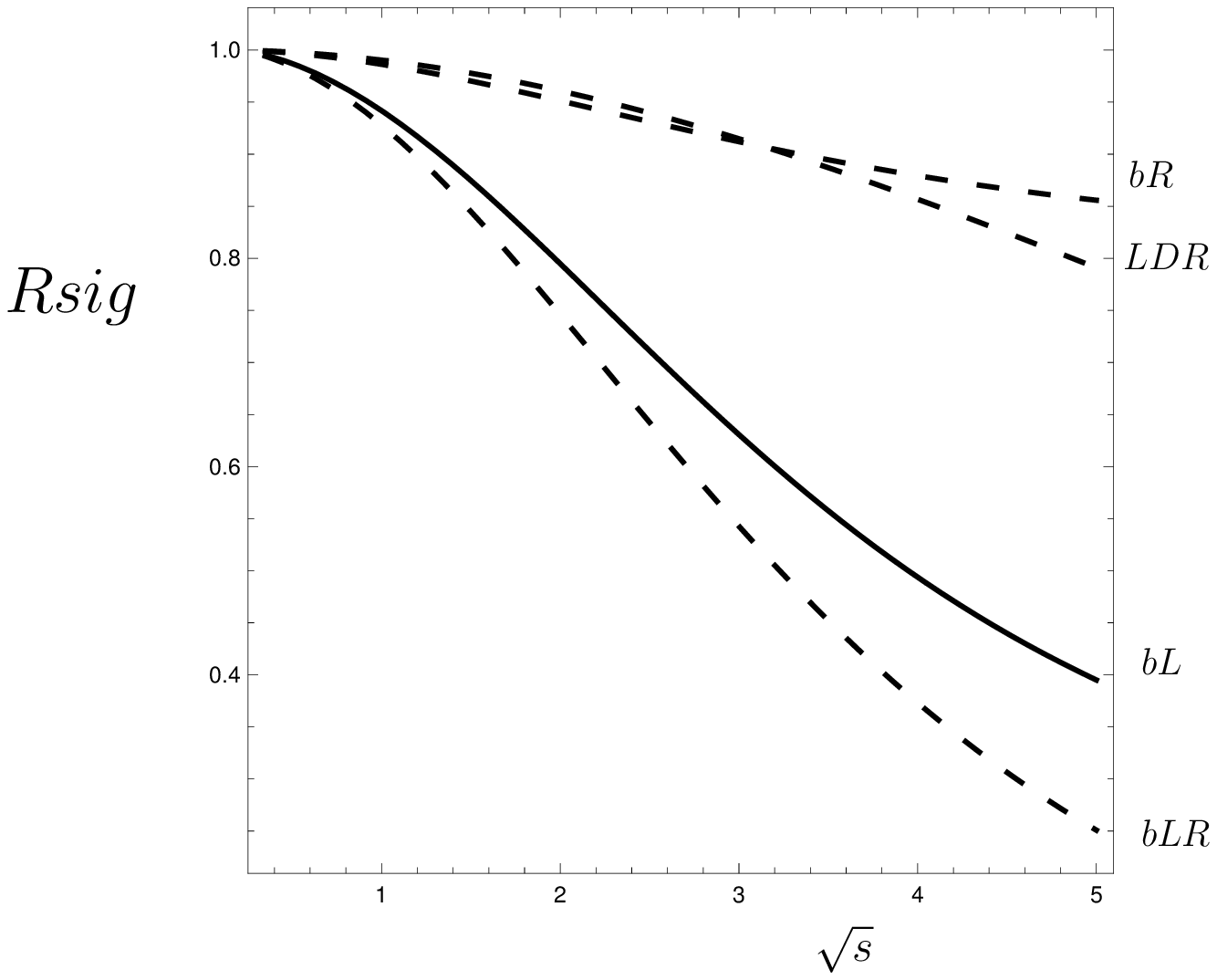, height=6.cm}
\hspace{1cm}
\epsfig{file=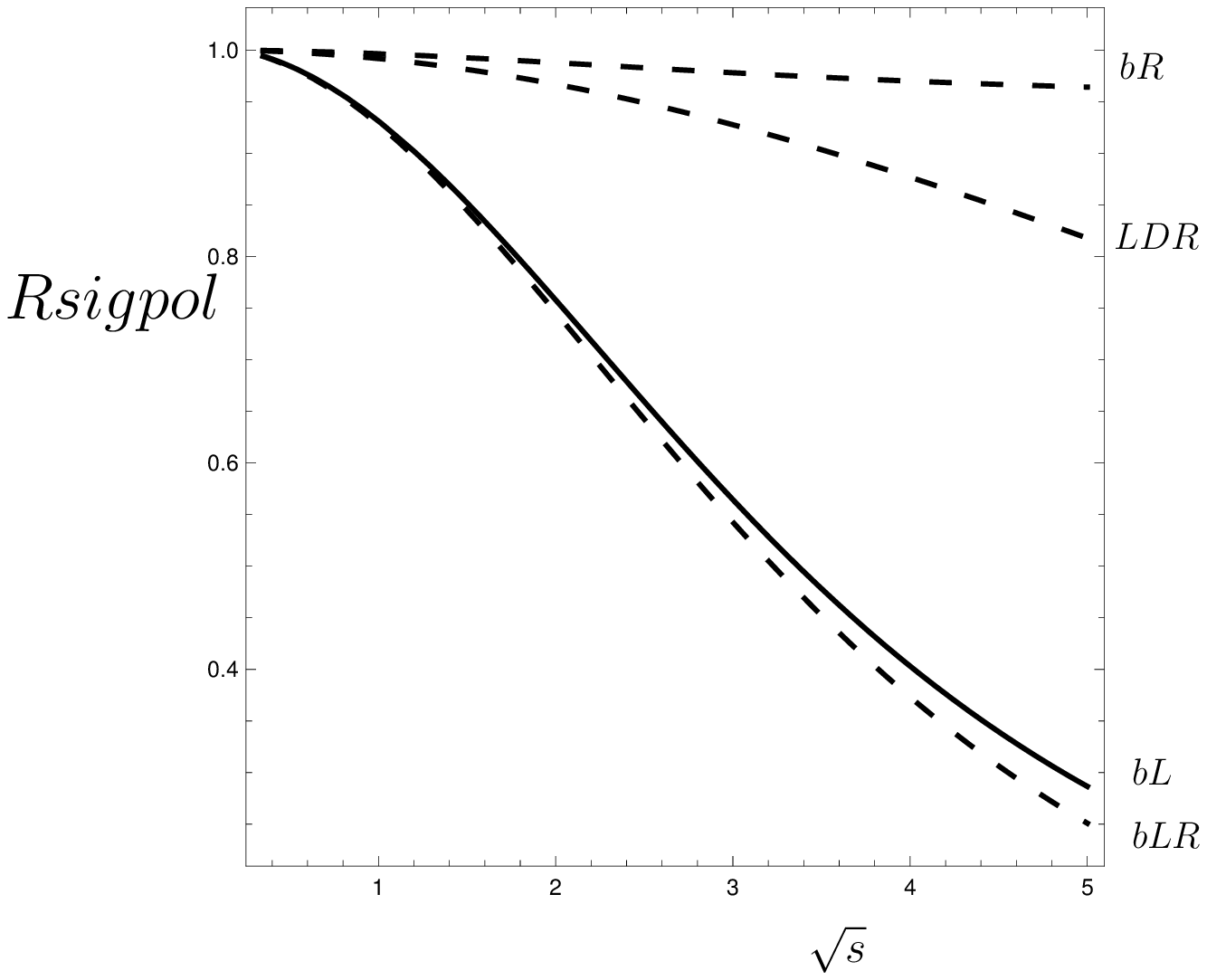, height=6.cm}
\]\\

\[
\epsfig{file=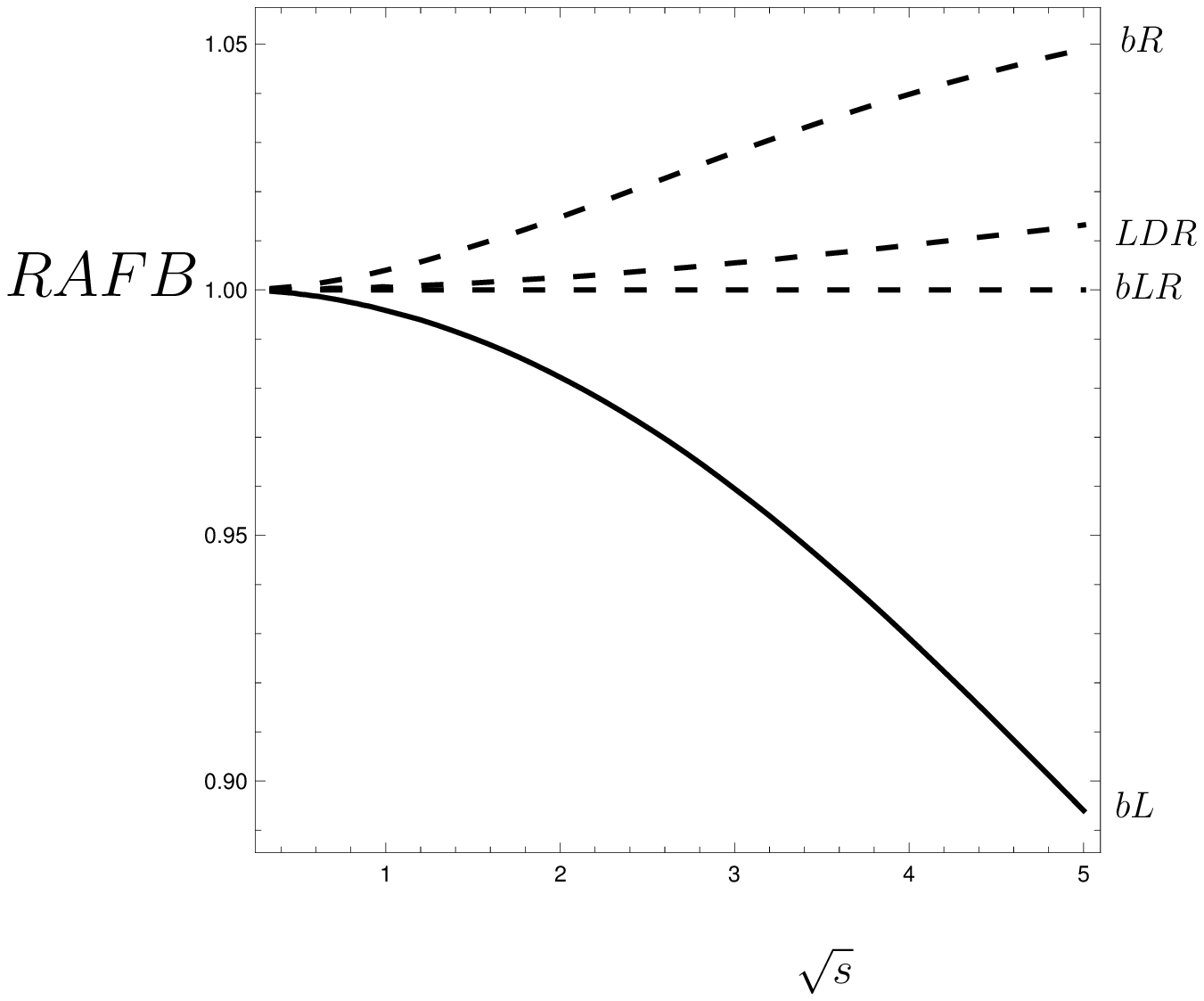, height=6.cm}
\hspace{1cm}
\epsfig{file=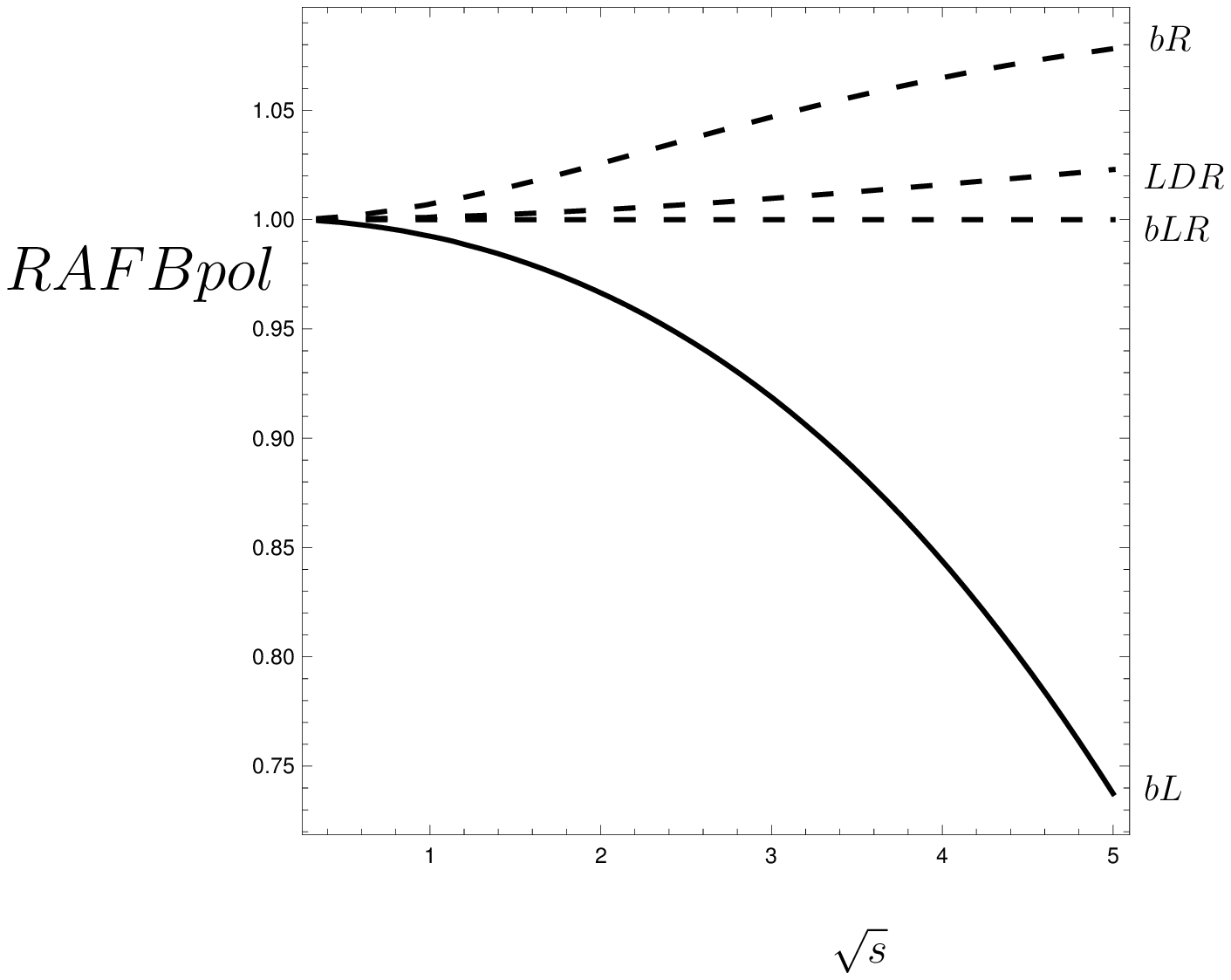, height=6.cm}
\]\\
\[
\epsfig{file=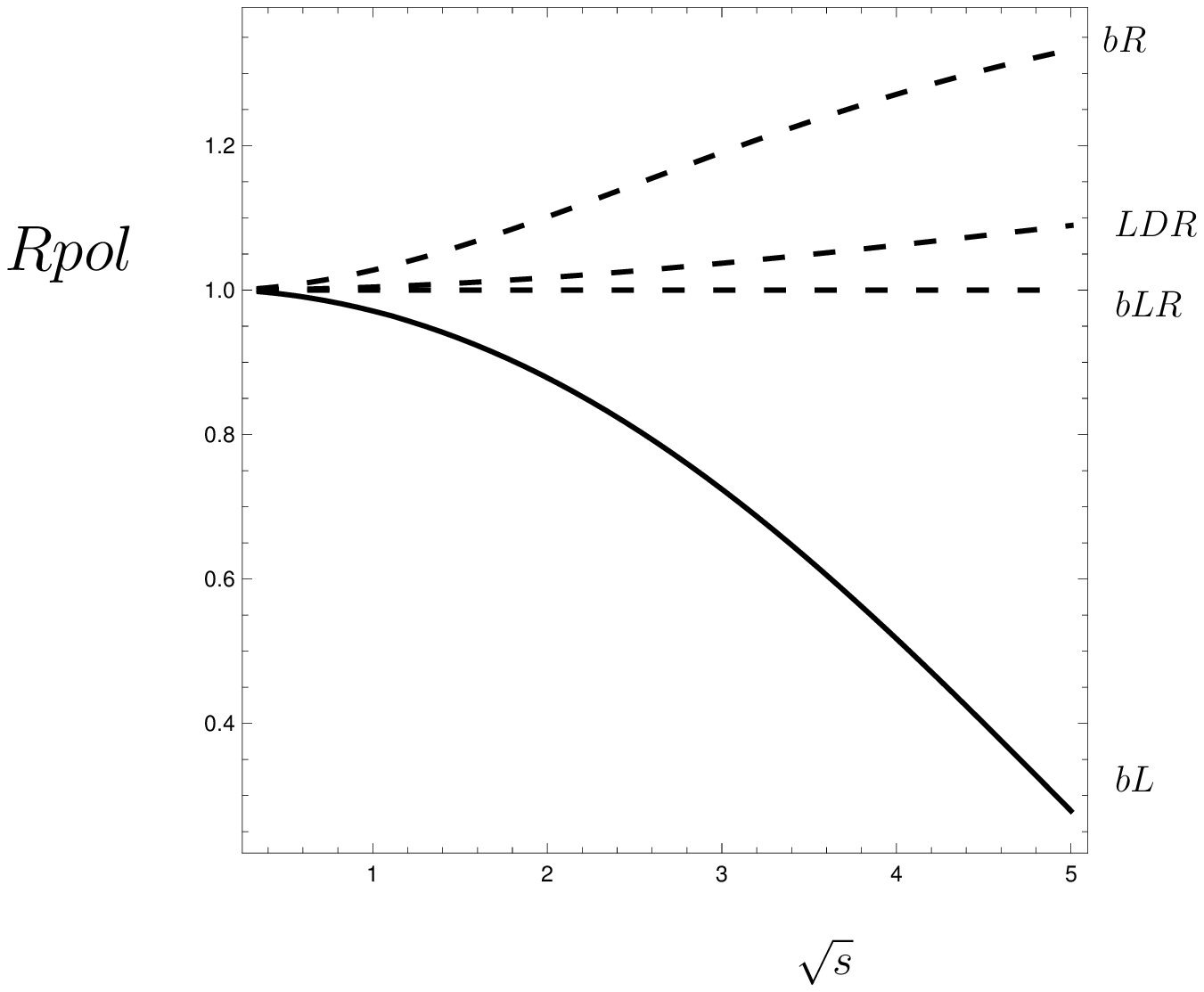, height=6.cm}
\hspace{1cm}
\epsfig{file=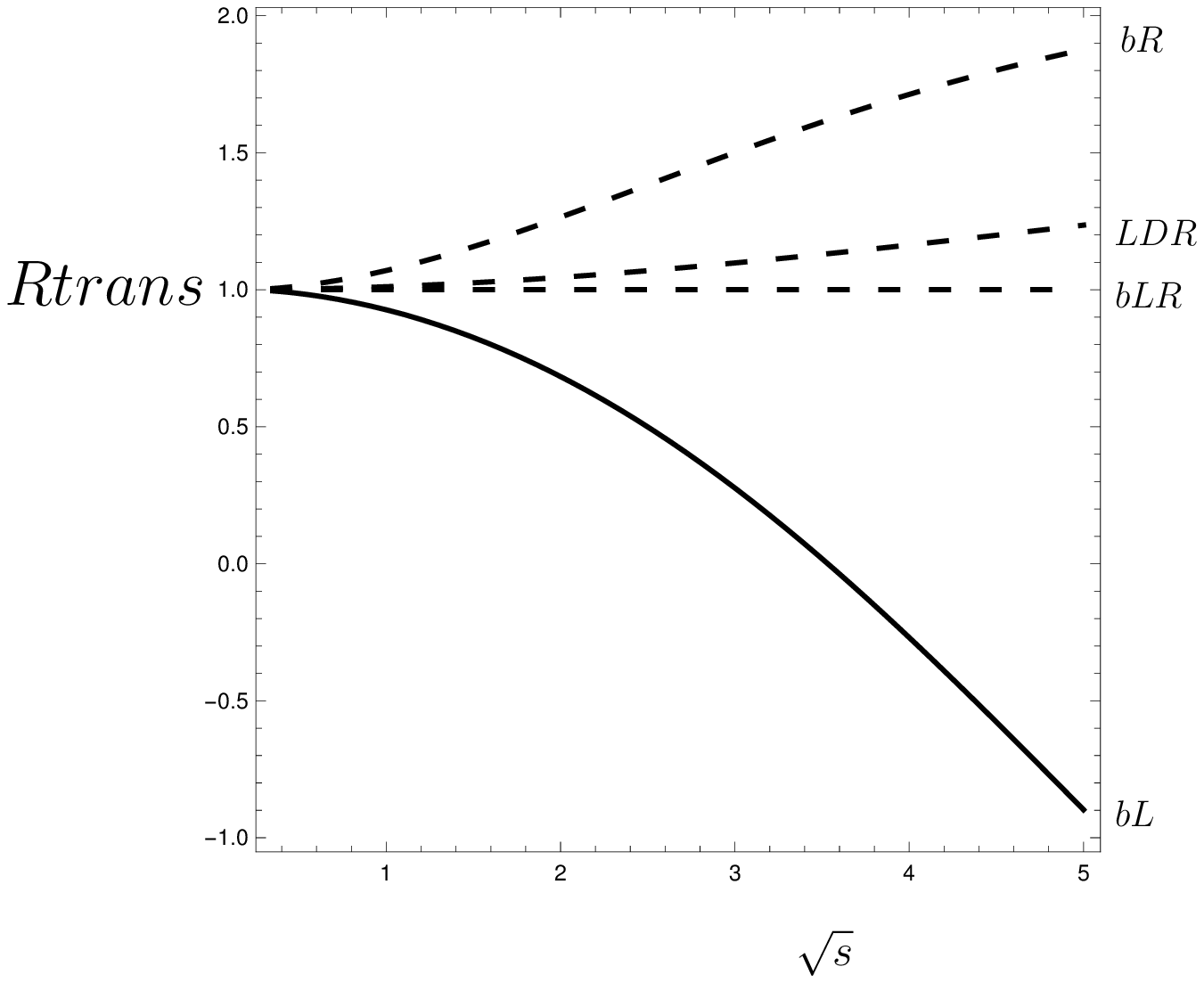, height=6.cm}
\]\\

\caption[1] {Ratios of $e^+e^-\to b\bar b$ observables with form factors
over the SM ones.}
\end{figure}

\clearpage

\begin{figure}[p]
\[
\epsfig{file=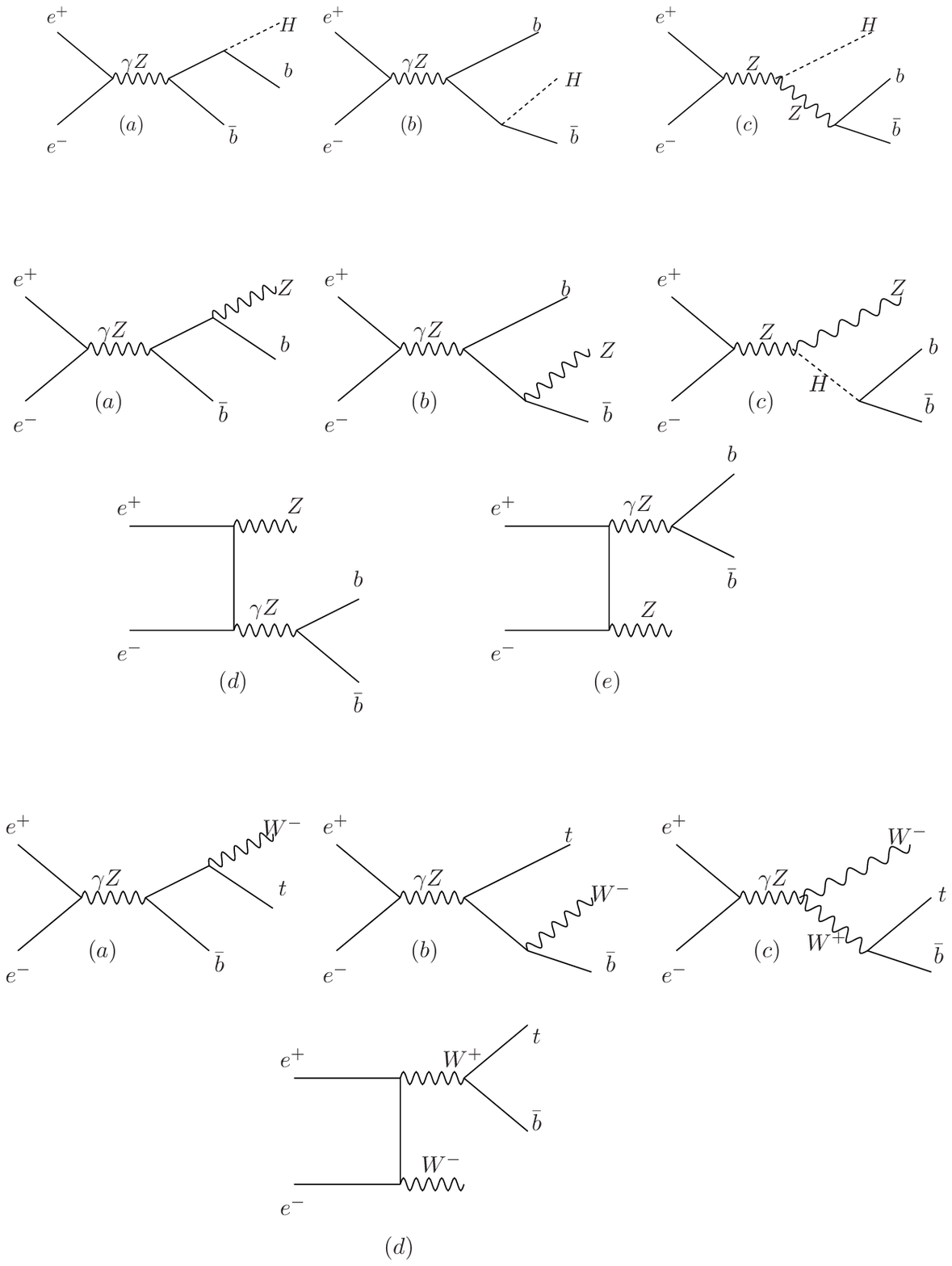, height=16.cm}
\]\\
\vspace{-1cm}
\caption[1]  {Diagrams for $e^+e^-\to b\bar b H, b\bar b Z, t\bar b W^-$.}

\end{figure}

\clearpage
\begin{figure}[p]
\[
\epsfig{file=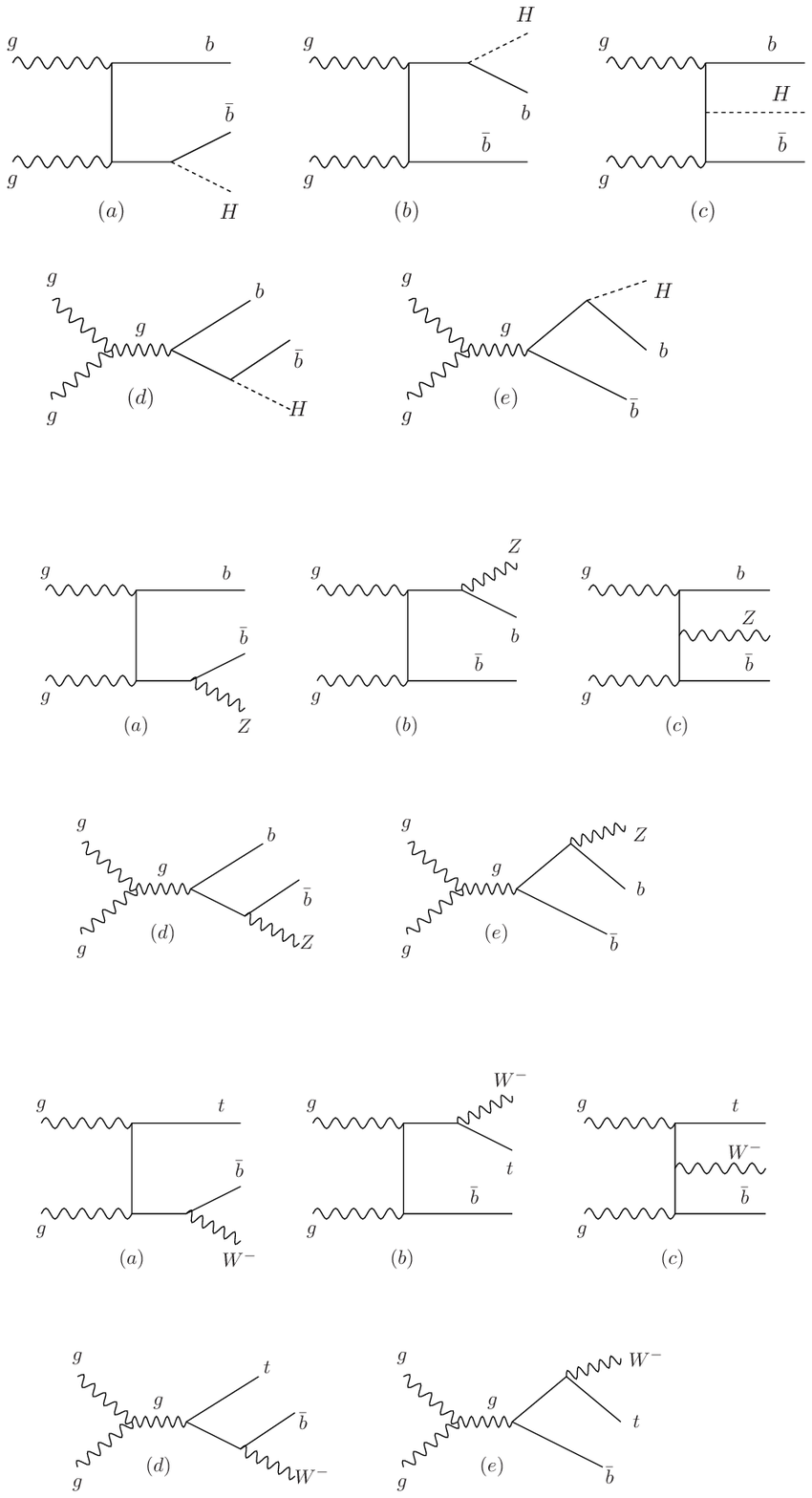, height=16.cm}
\]\\
\vspace{-1cm}
\caption[1] {Diagrams for $gg\to b\bar b H, b\bar b Z, t\bar b W^-$.}
\clearpage
\end{figure}
\begin{figure}[p]
\[
\epsfig{file=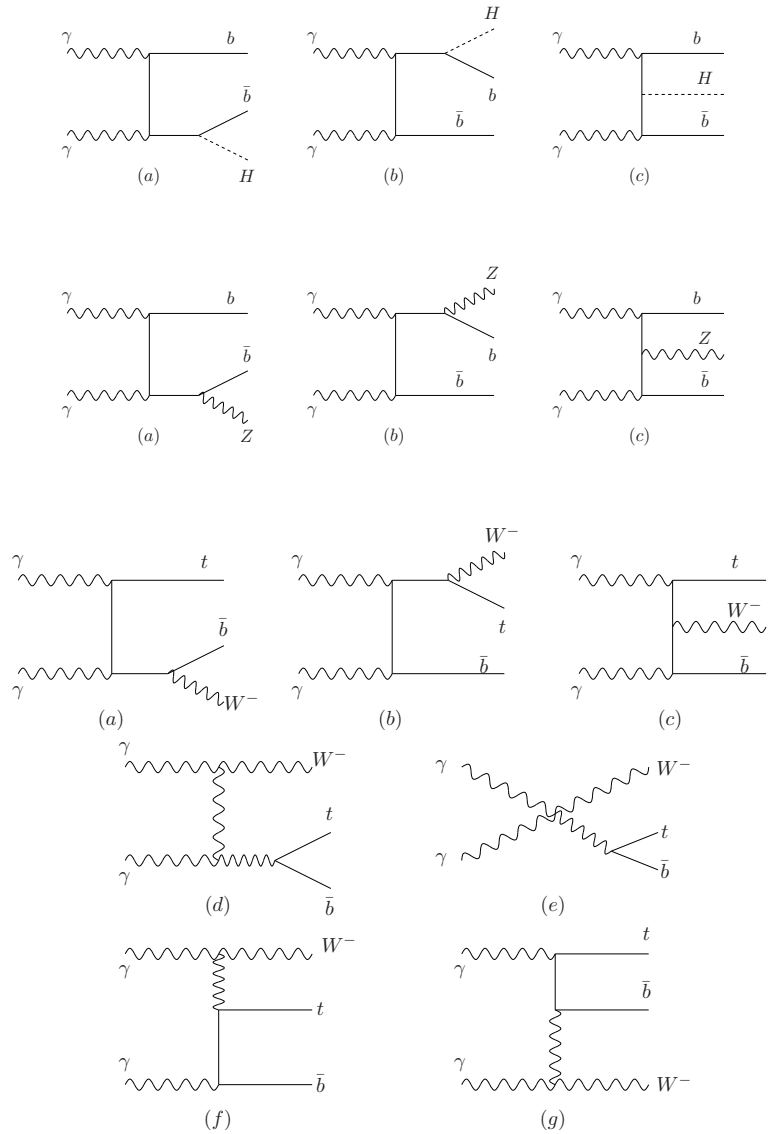, height=24.cm}
\]\\
\vspace{-6cm}
\caption[1] {Diagrams for $\gamma\gamma\to b\bar b H, b\bar b Z, t\bar b W^-$.}

\end{figure}
    
\clearpage

\begin{figure}[p]
\[
\epsfig{file=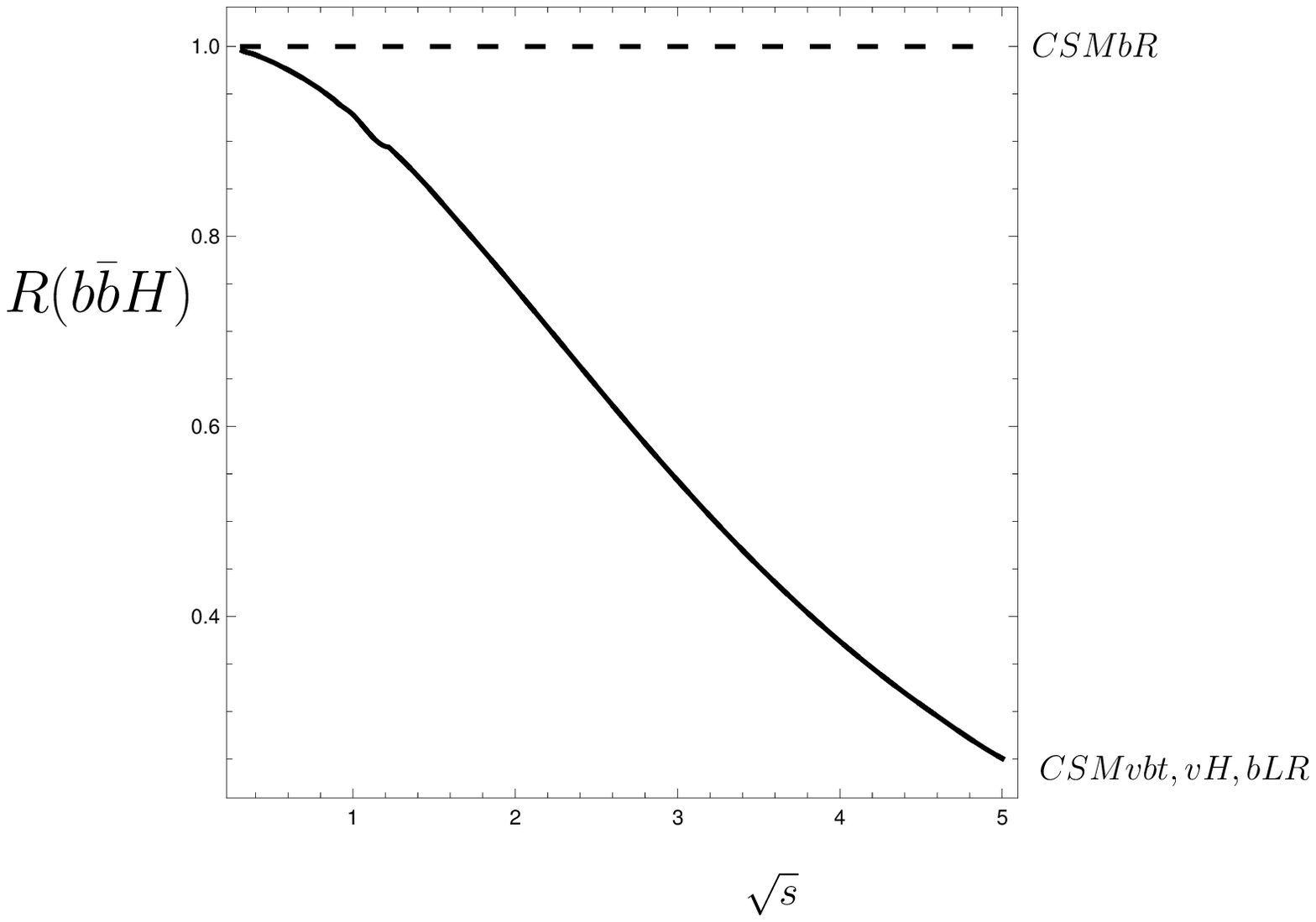, height=8.cm}
\]\\
\vspace{-1cm}
\[
\epsfig{file=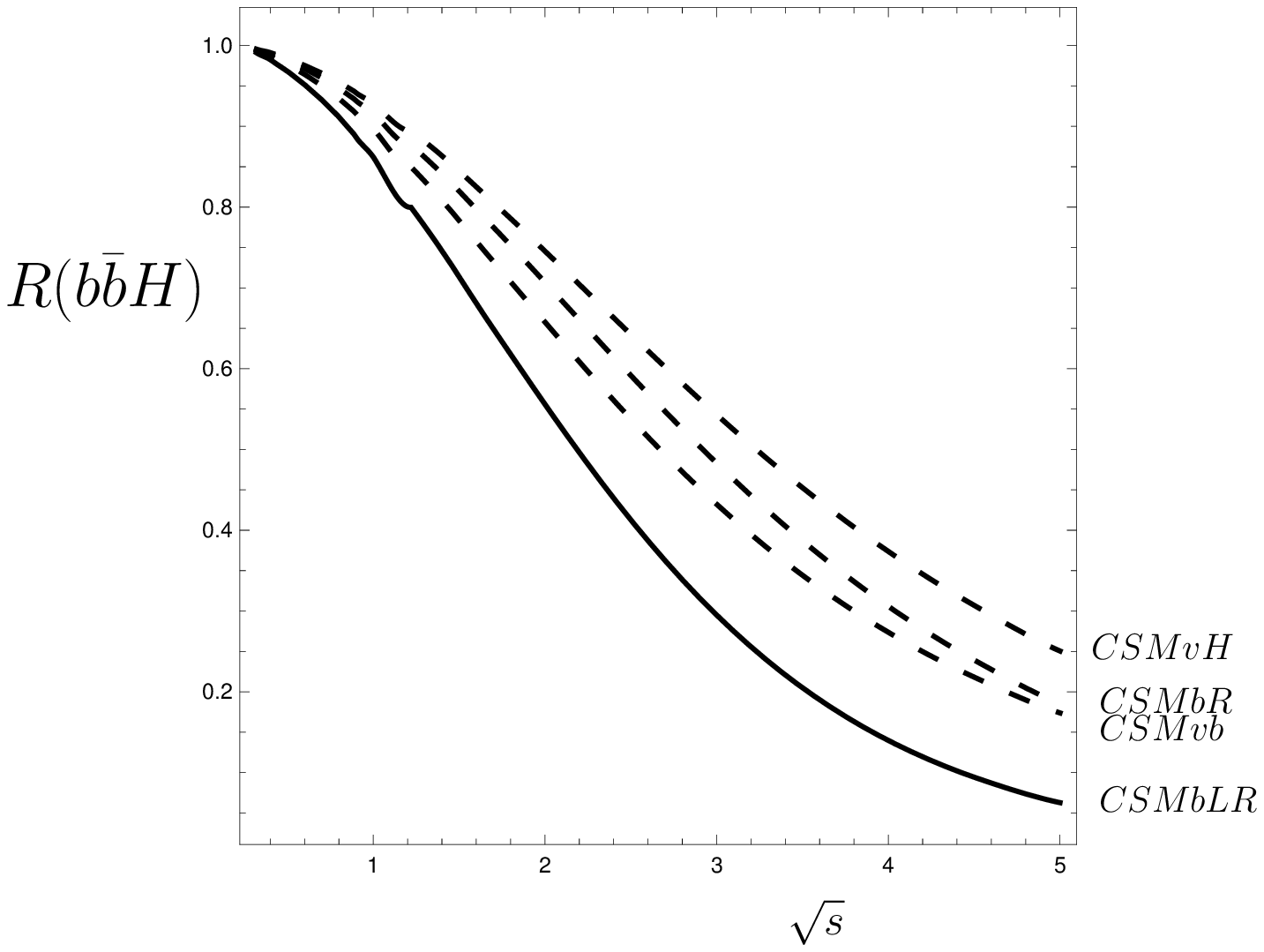, height=8.cm}
\]\\
\caption[1]  {Ratio of $e^+e^-\to b\bar b H$ cross section for elementary $b$ 
with only $H$ form factor (up) and  with all form factors for full composite $b$ (down) over the standard one.}
\end{figure}

\clearpage

\begin{figure}[p]
\[
\epsfig{file=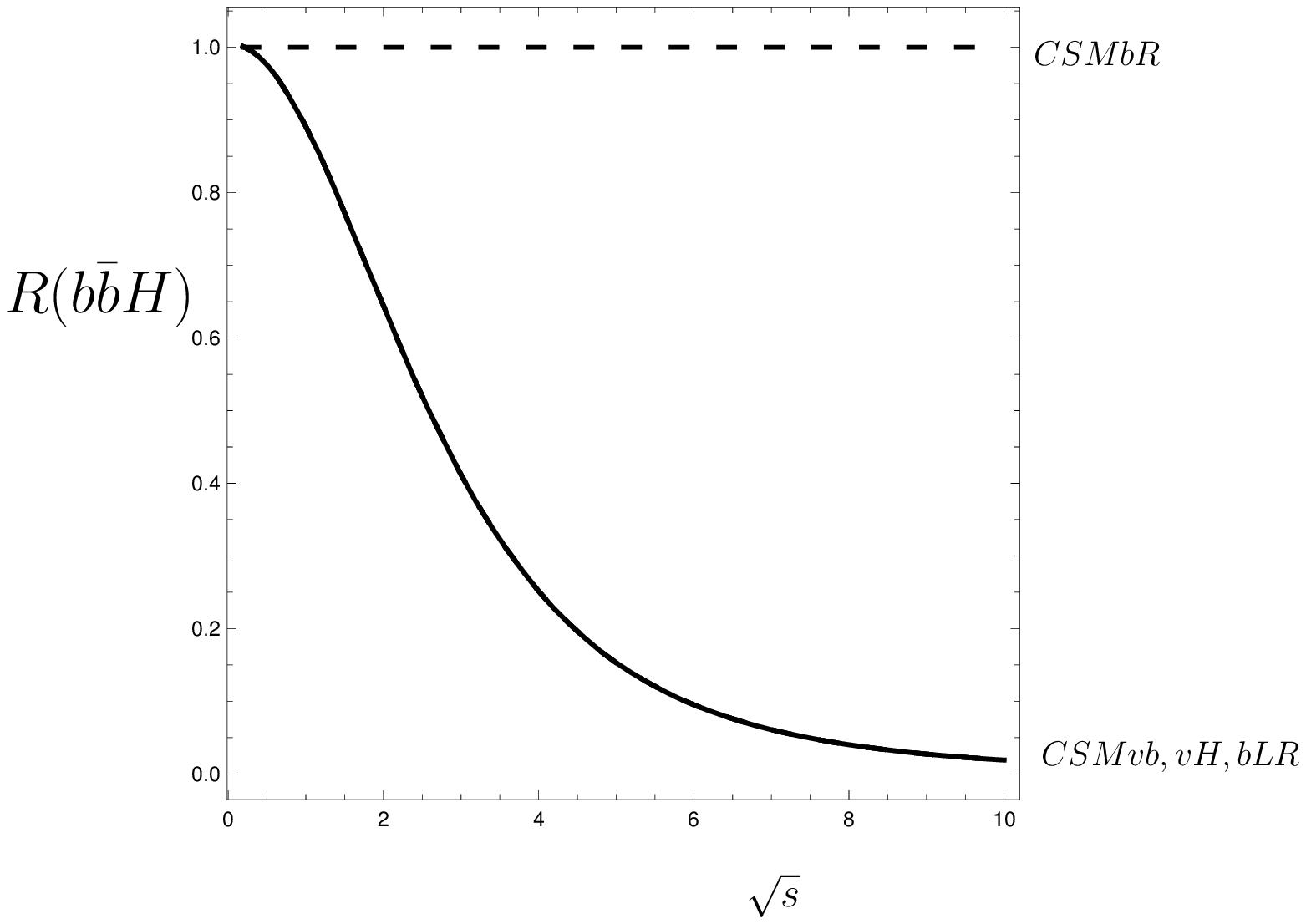, height=8.cm}
\]\\
\vspace{-1cm}
\[
\epsfig{file=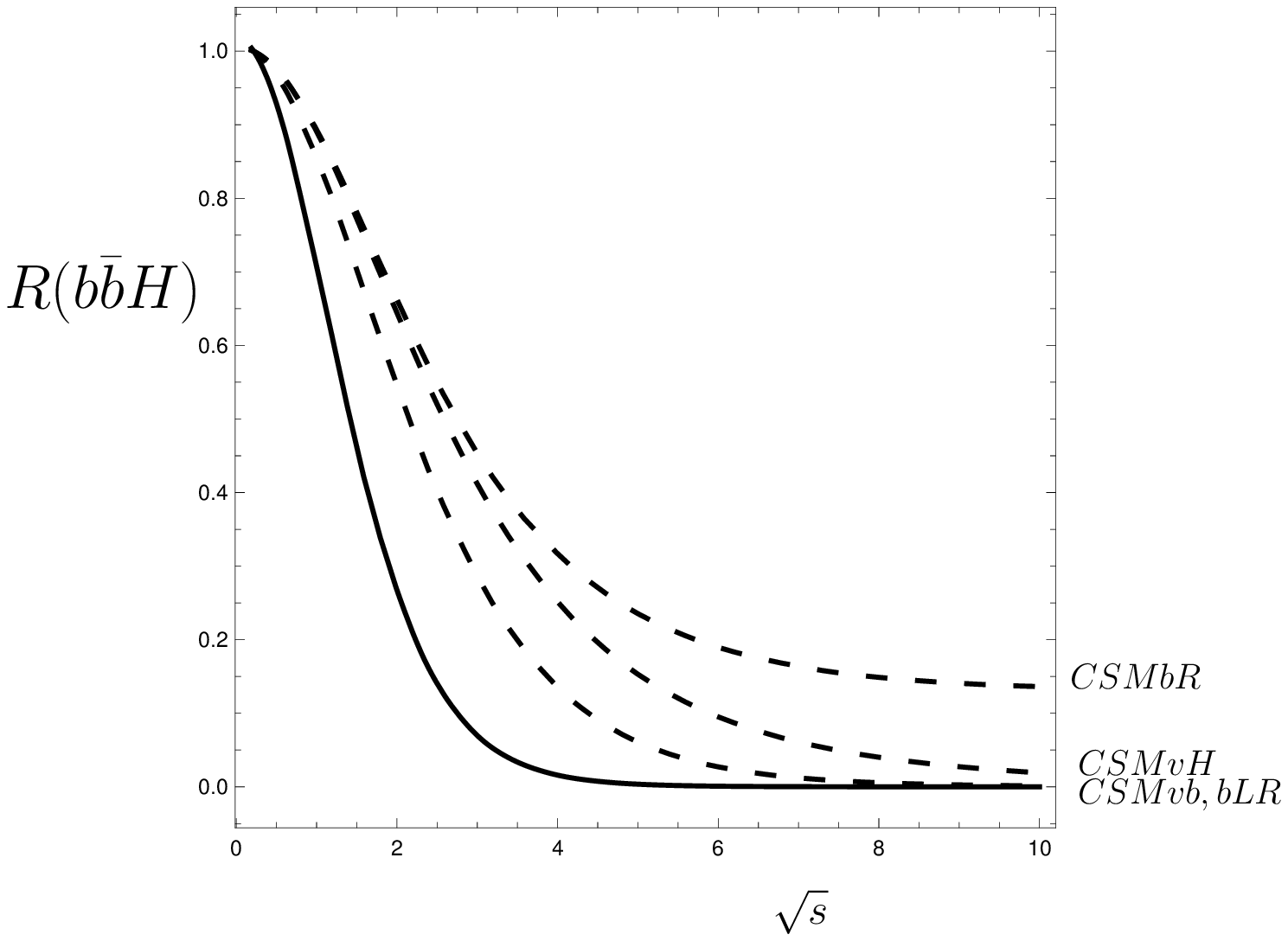, height=8.cm}
\]\\
\caption[1]  {Ratio of $\gamma \gamma \to b\bar b H$ cross section for elementary $b$
with only $H$ form factor (up) and  with all form factors for full composite $b$ (down) over the standard one.}
\end{figure}

\clearpage

\begin{figure}[p]
\[
\epsfig{file=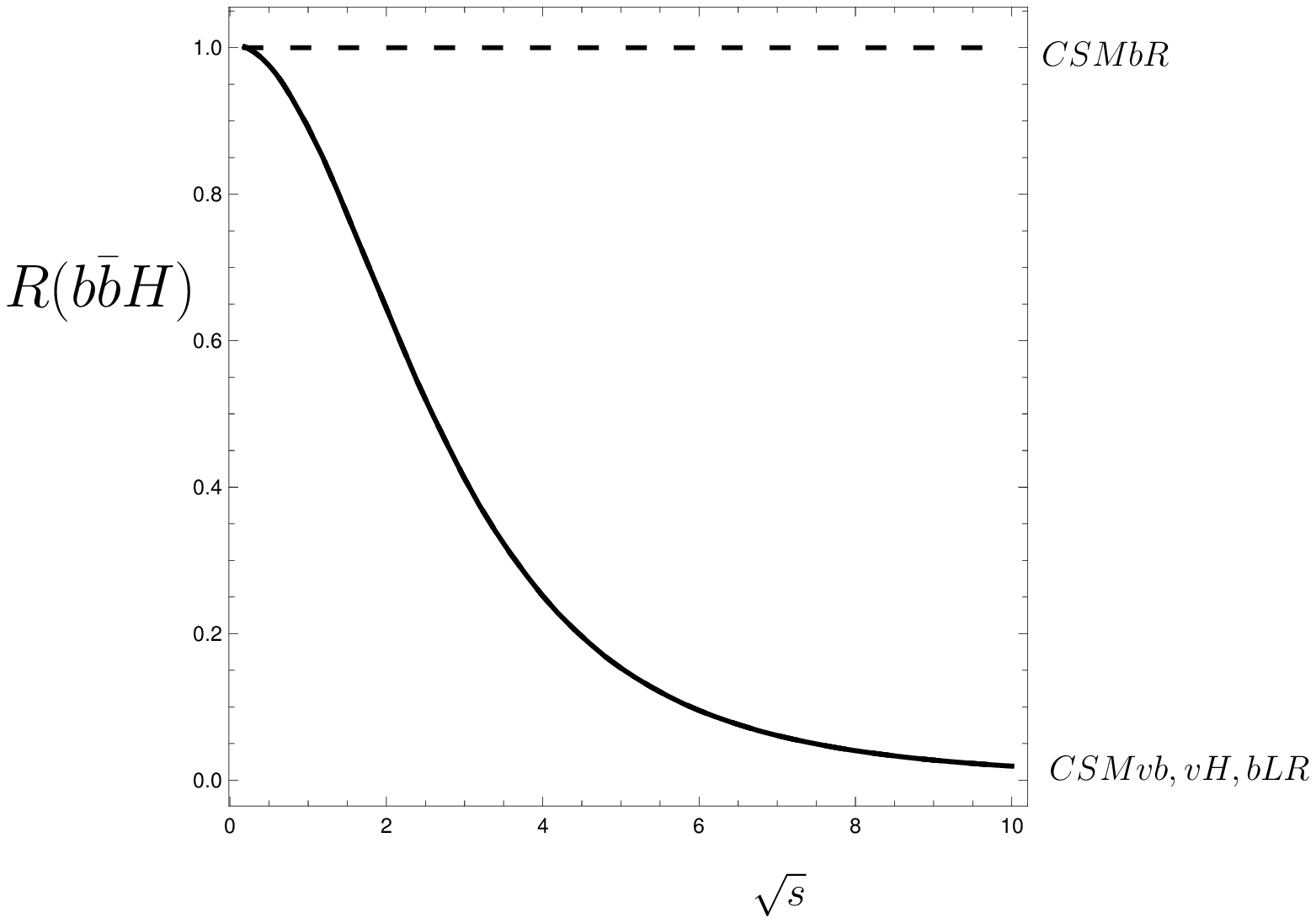, height=8.cm}
\]\\
\vspace{-1cm}
\[
\epsfig{file=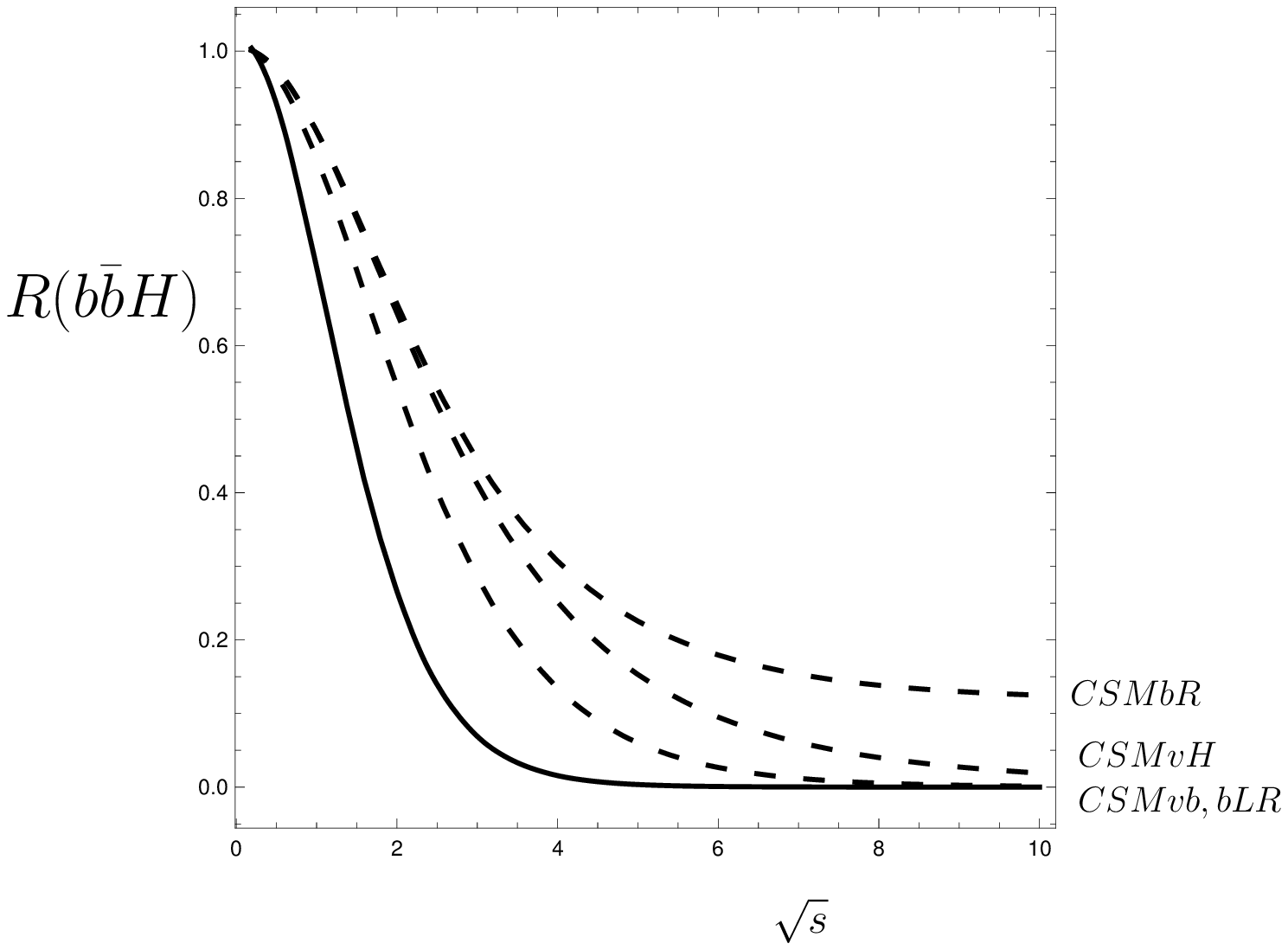, height=8.cm}
\]\\
\caption[1] {Ratio of $gg\to b\bar b H$ cross section for elementary $b$
with only $H$ form factor (up) and with all form factors for full composite $b$ (down) over the standard one.}
\end{figure}

\clearpage

\begin{figure}[p]
\[
\epsfig{file=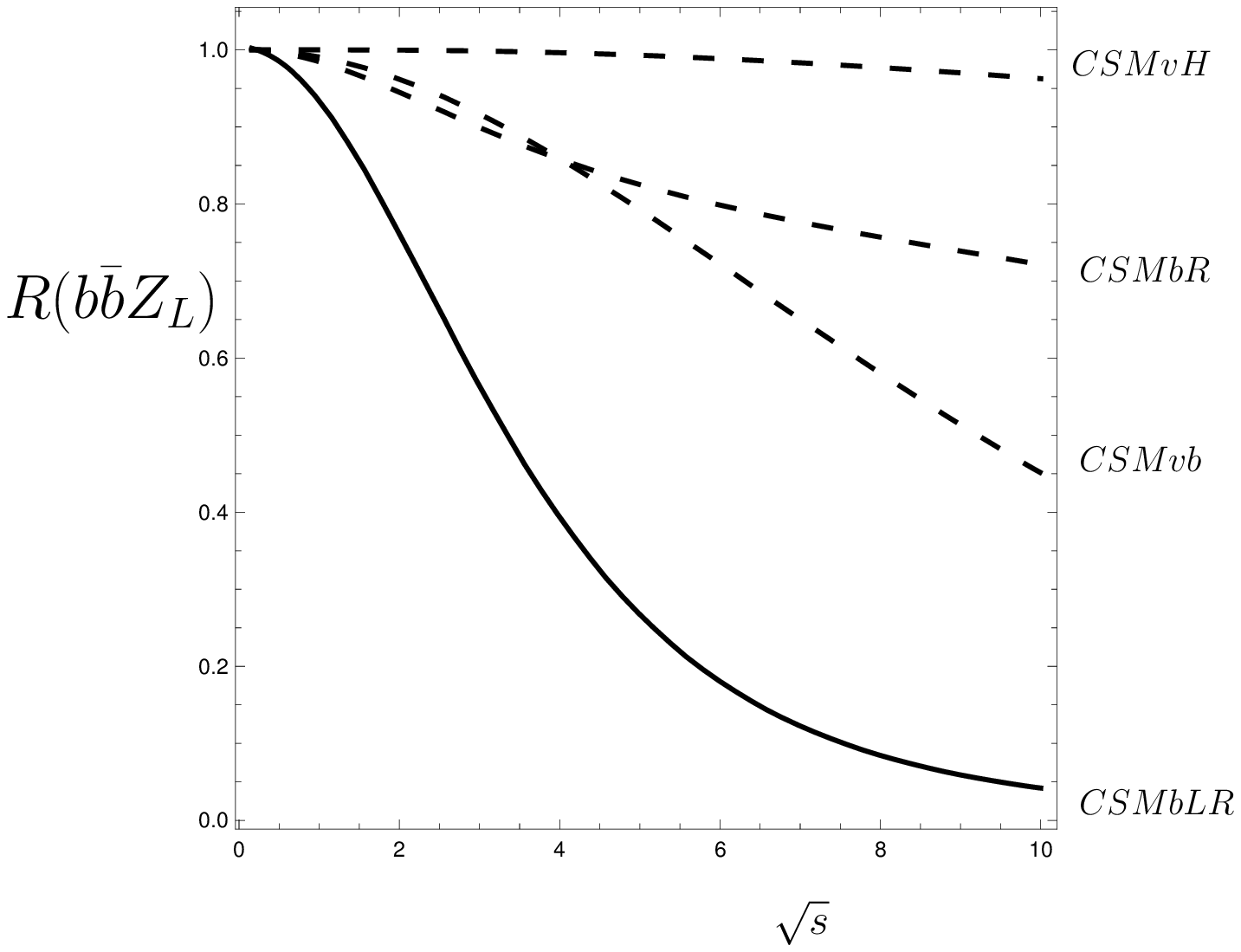, height=6.cm}
\]\\
\vspace{-1cm}
\[
\epsfig{file=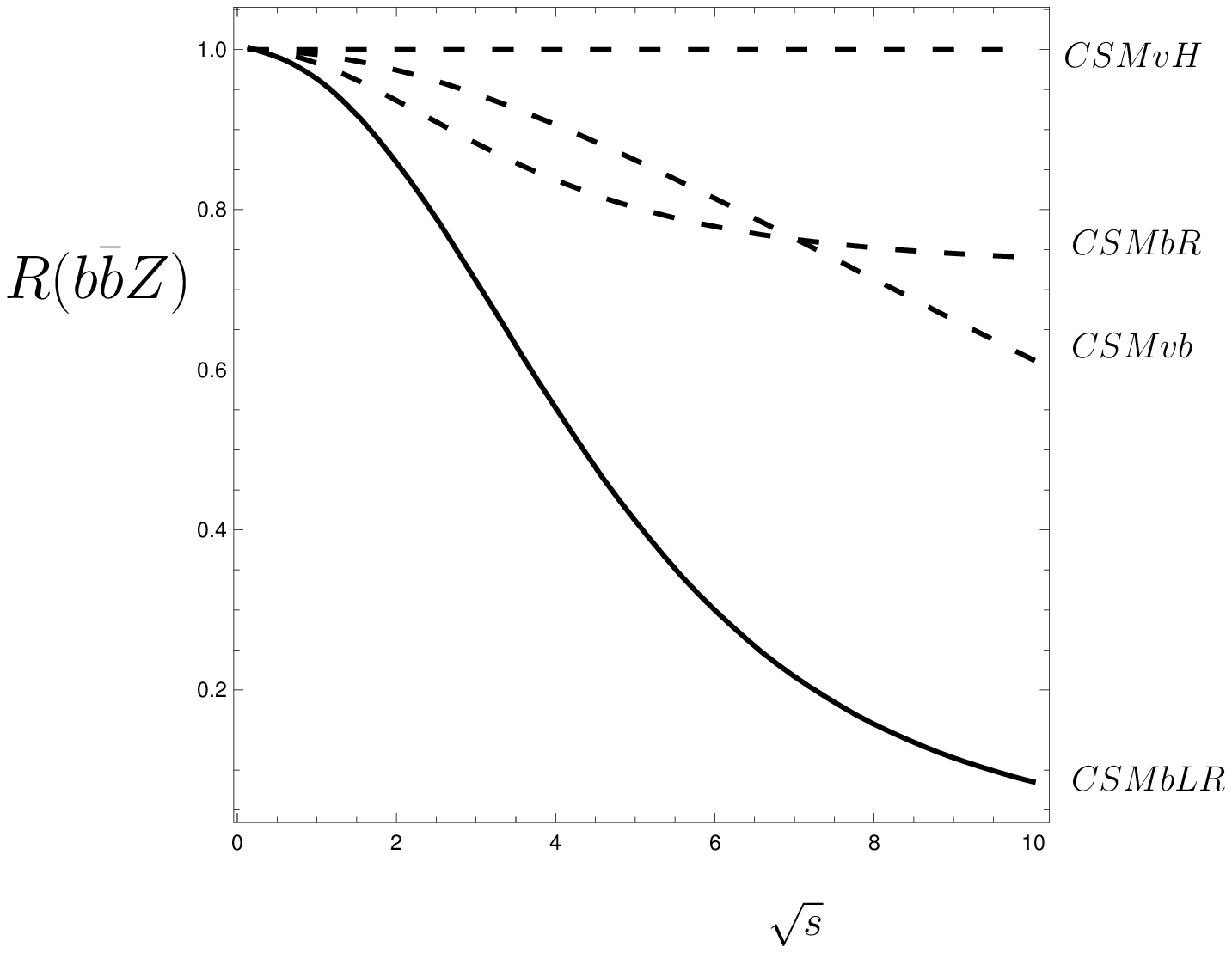, height=6.cm}
\]\\
\caption[1] {Ratio of $e^+e^-\to b\bar b Z_L$ (up) and unpolarized $Z$ (down)
cross section with form factors (for full composite $b$)
over the standard one.}
\end{figure}

\clearpage

\begin{figure}[p]
\[
\epsfig{file=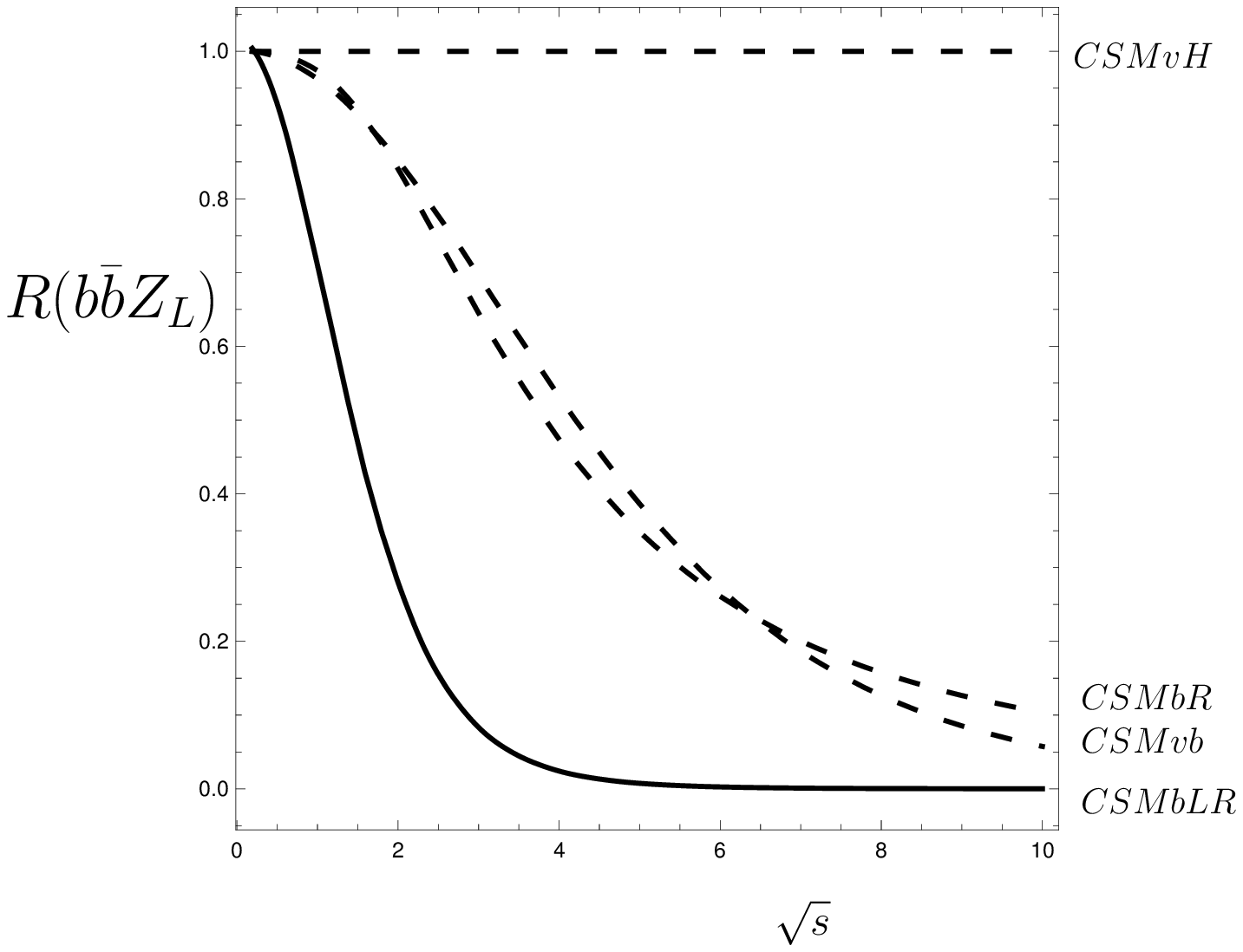, height=8.cm}
\]\\
\vspace{-1cm}
\[
\epsfig{file=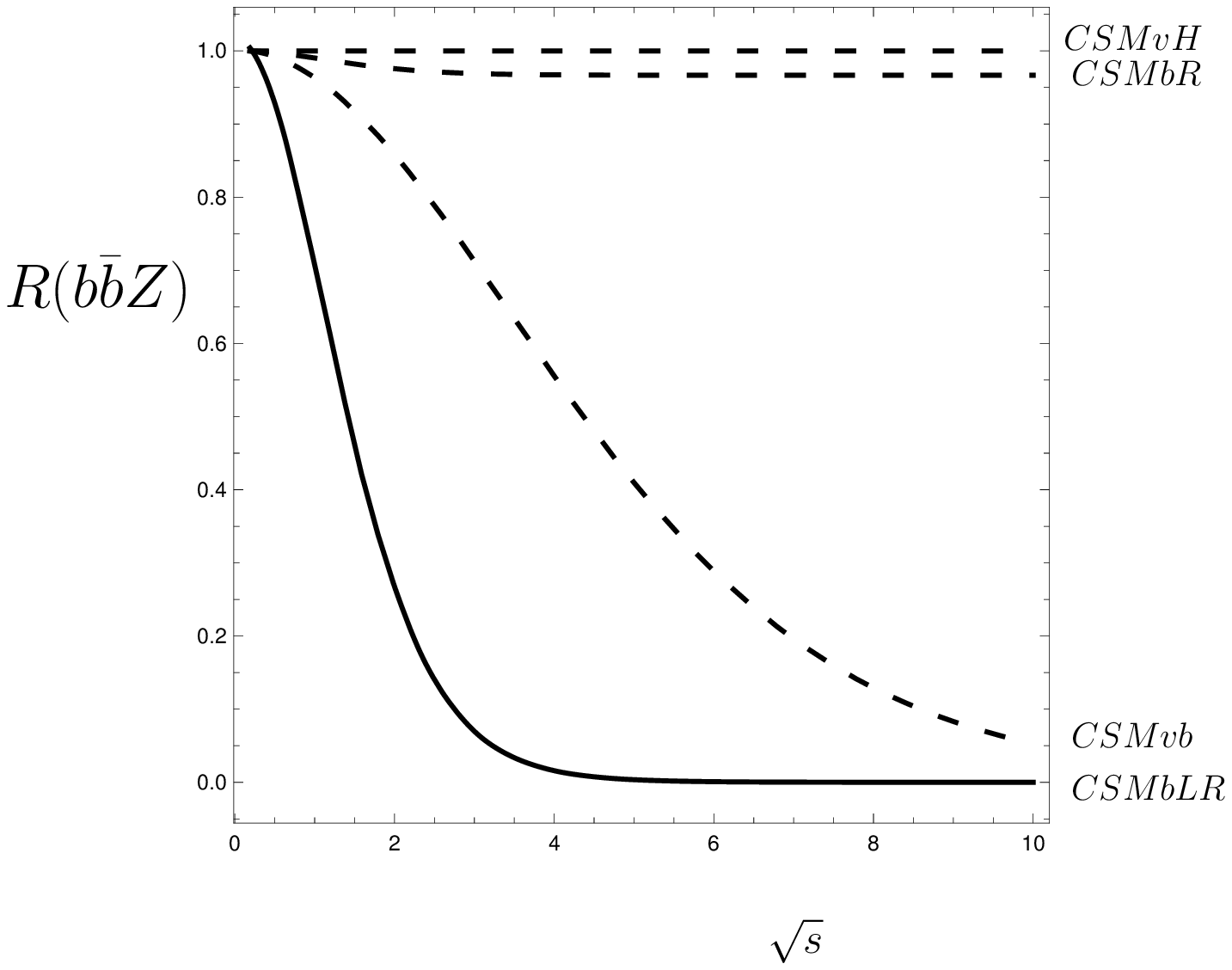, height=8.cm}
\]\\
\caption[1] {Ratio of $\gamma\gamma\to b\bar b Z_L$ (up) and unpolarized $Z$ (down)
cross section with form factors
(for full composite $b$) over the standard one.}
\end{figure}

\clearpage

\begin{figure}[p]
\[
\epsfig{file=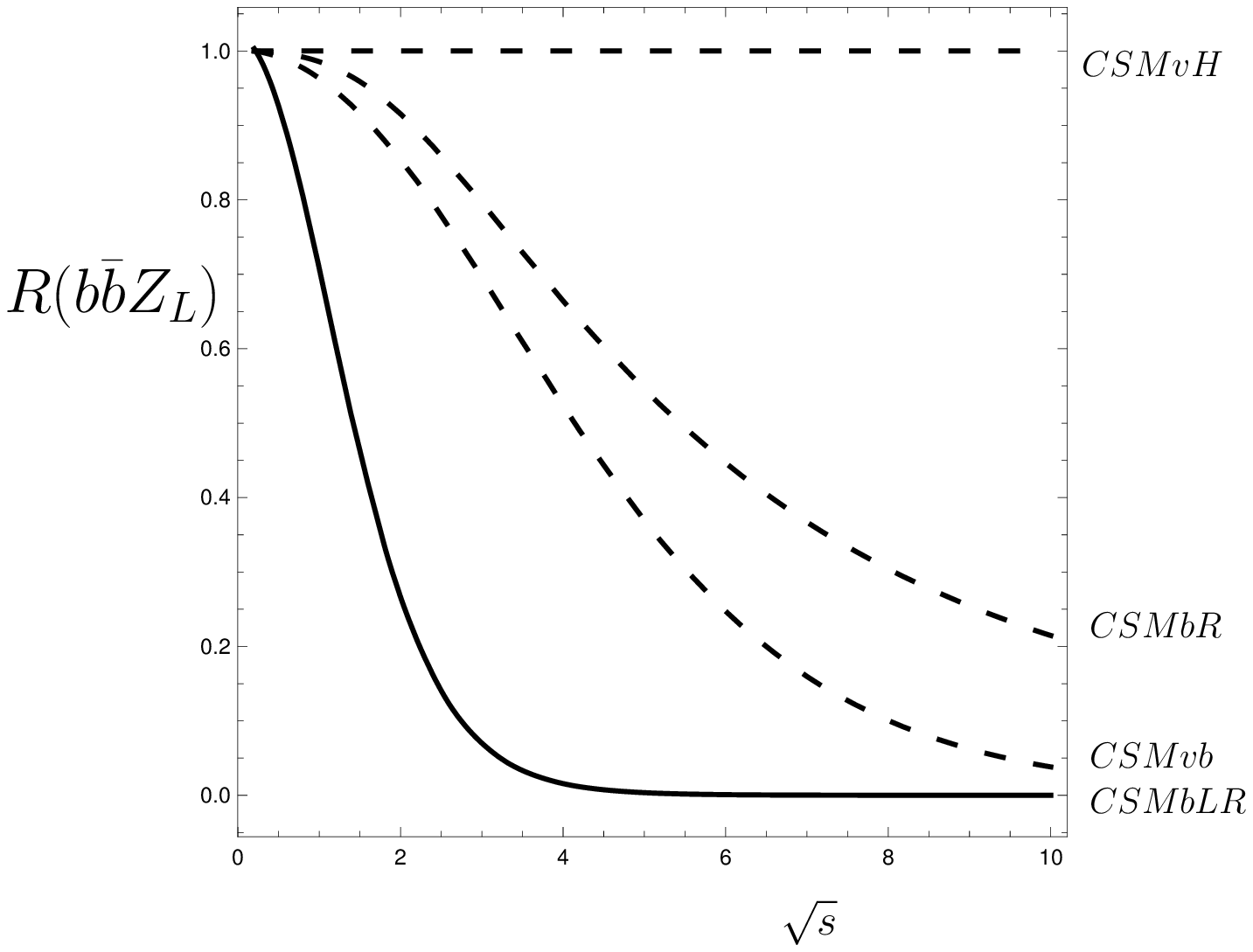, height=8.cm}
\]\\
\vspace{-1cm}
\[
\epsfig{file=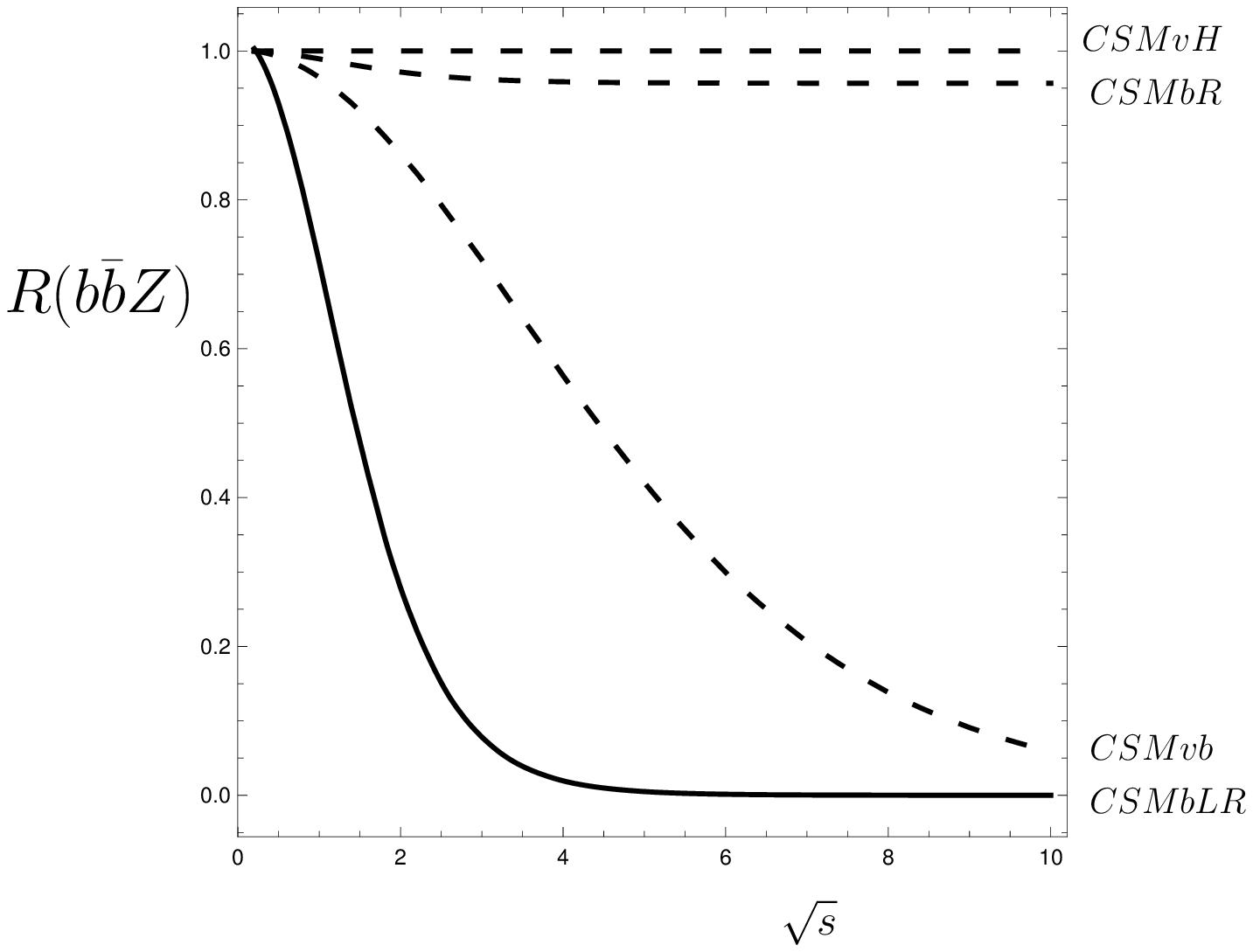, height=8.cm}
\]\\
\caption[1] {Ratio of $gg\to b\bar b Z_L$ (up) and unpolarized $Z$ (down)
cross section with form factors (for full composite $b$) over the standard one.}
\end{figure}

\clearpage

\begin{figure}[p]
\[
\epsfig{file=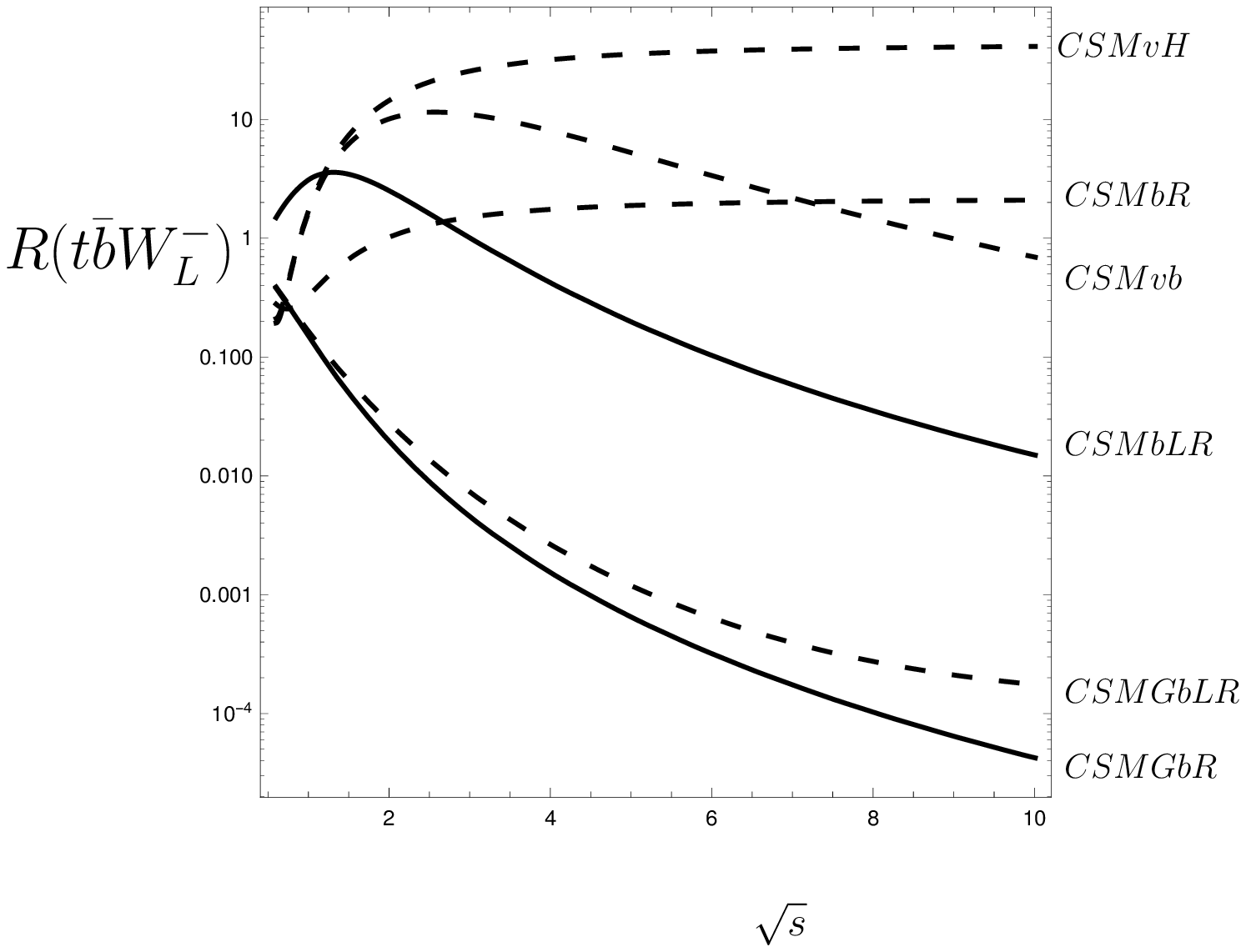, height=6.cm}
\epsfig{file=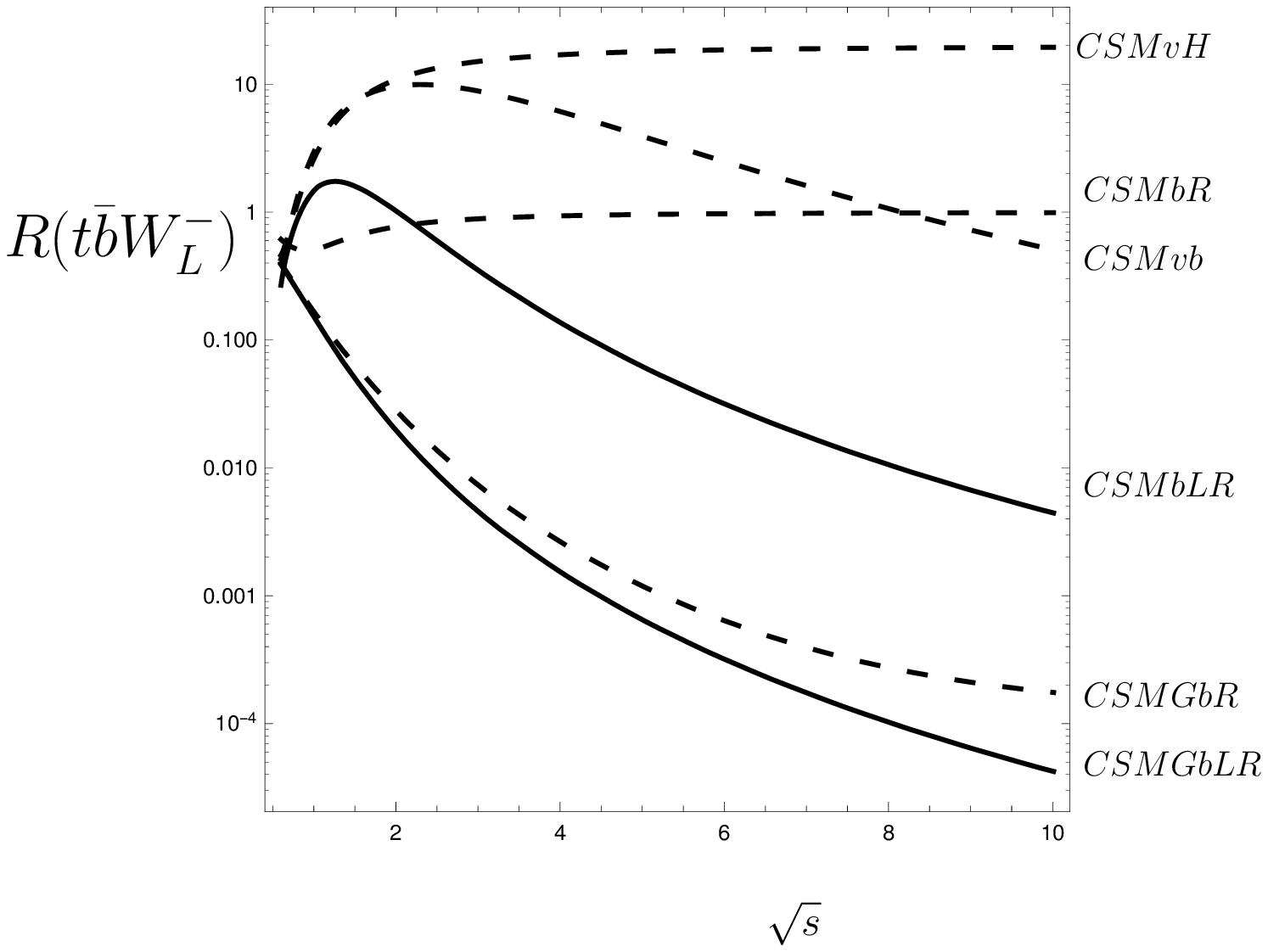, height=6.cm}
\]\\

\[
\epsfig{file=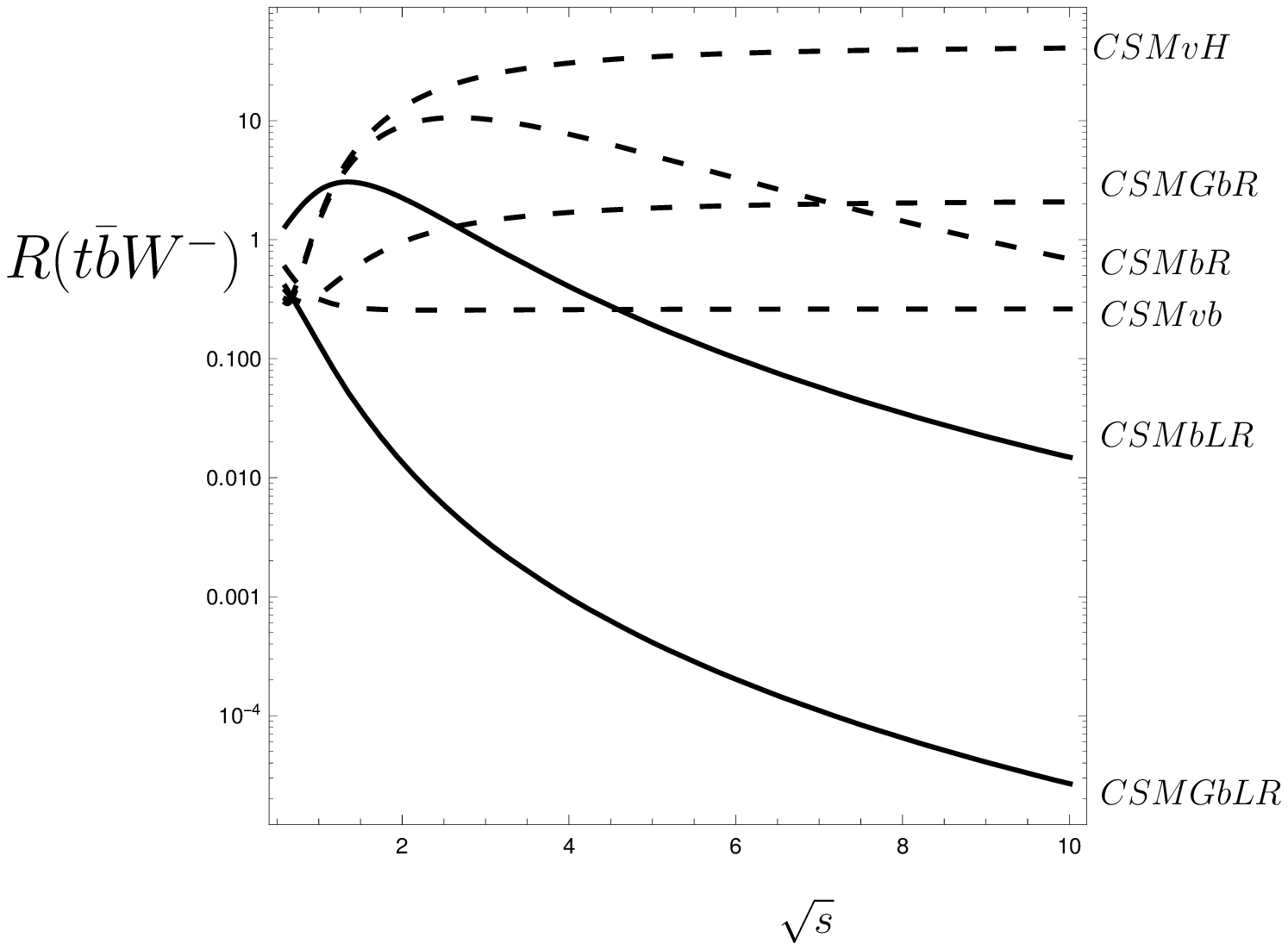, height=6.cm}
\epsfig{file=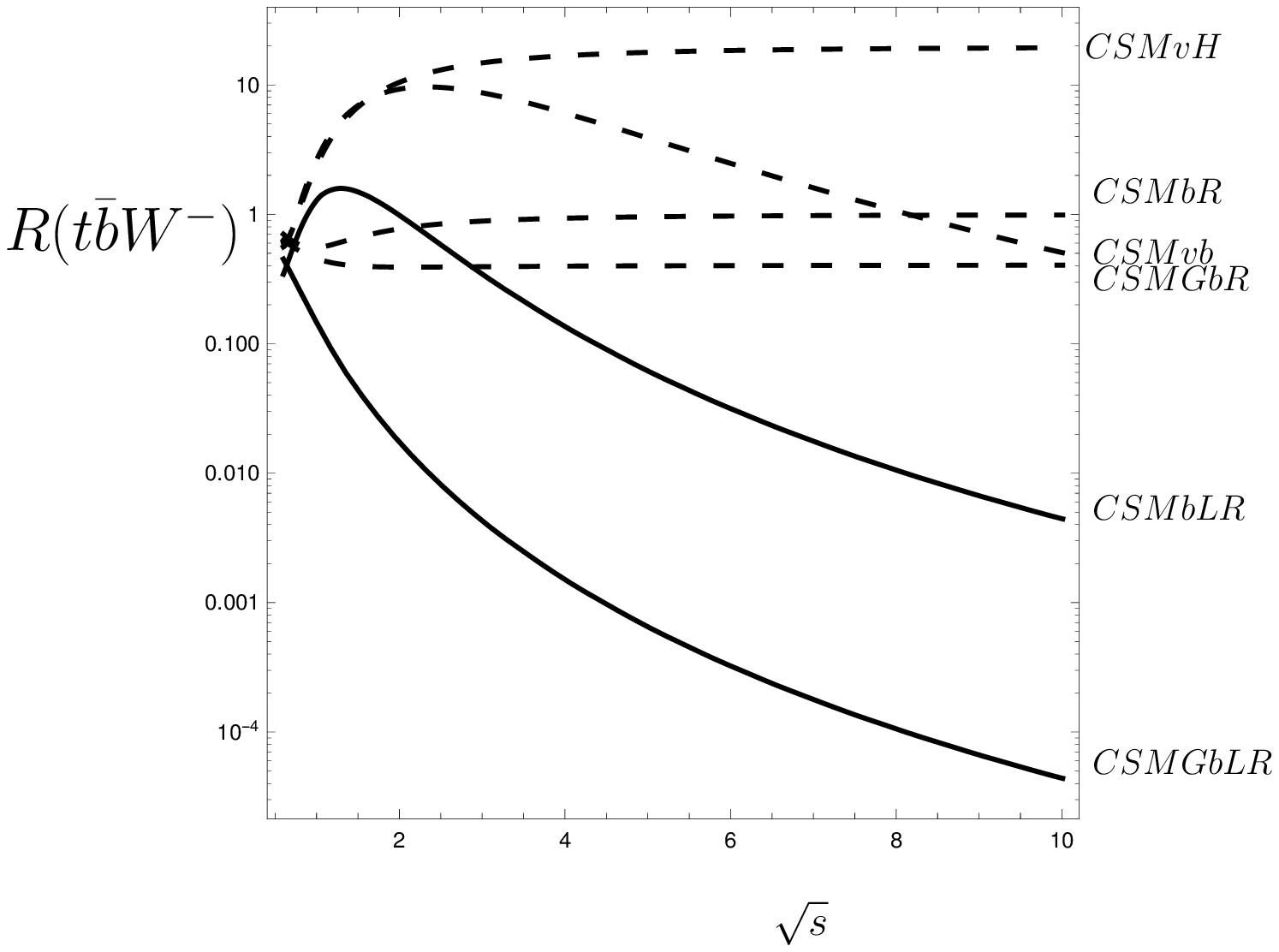, height=6.cm}
\]\\
\caption[1] {Same ratios for $e^+e^-\to t\bar b W^-_L$ 
and $e^+e^-\to t\bar b W^-$   for elementary $b$ (left),
for full composite $b$ (right).}
\end{figure}

\clearpage

\begin{figure}[p]
\[
\epsfig{file=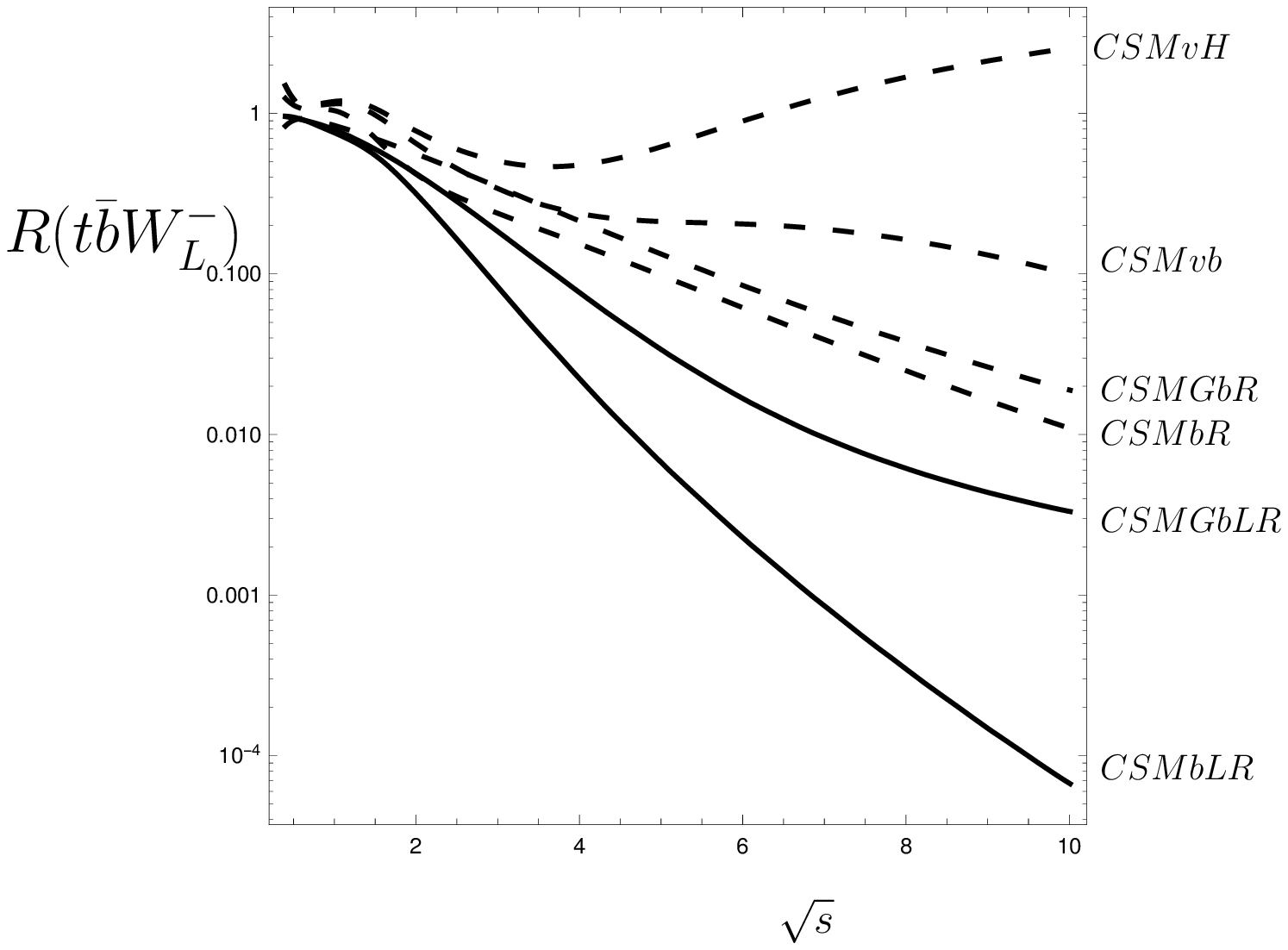, height=6.cm}
\epsfig{file=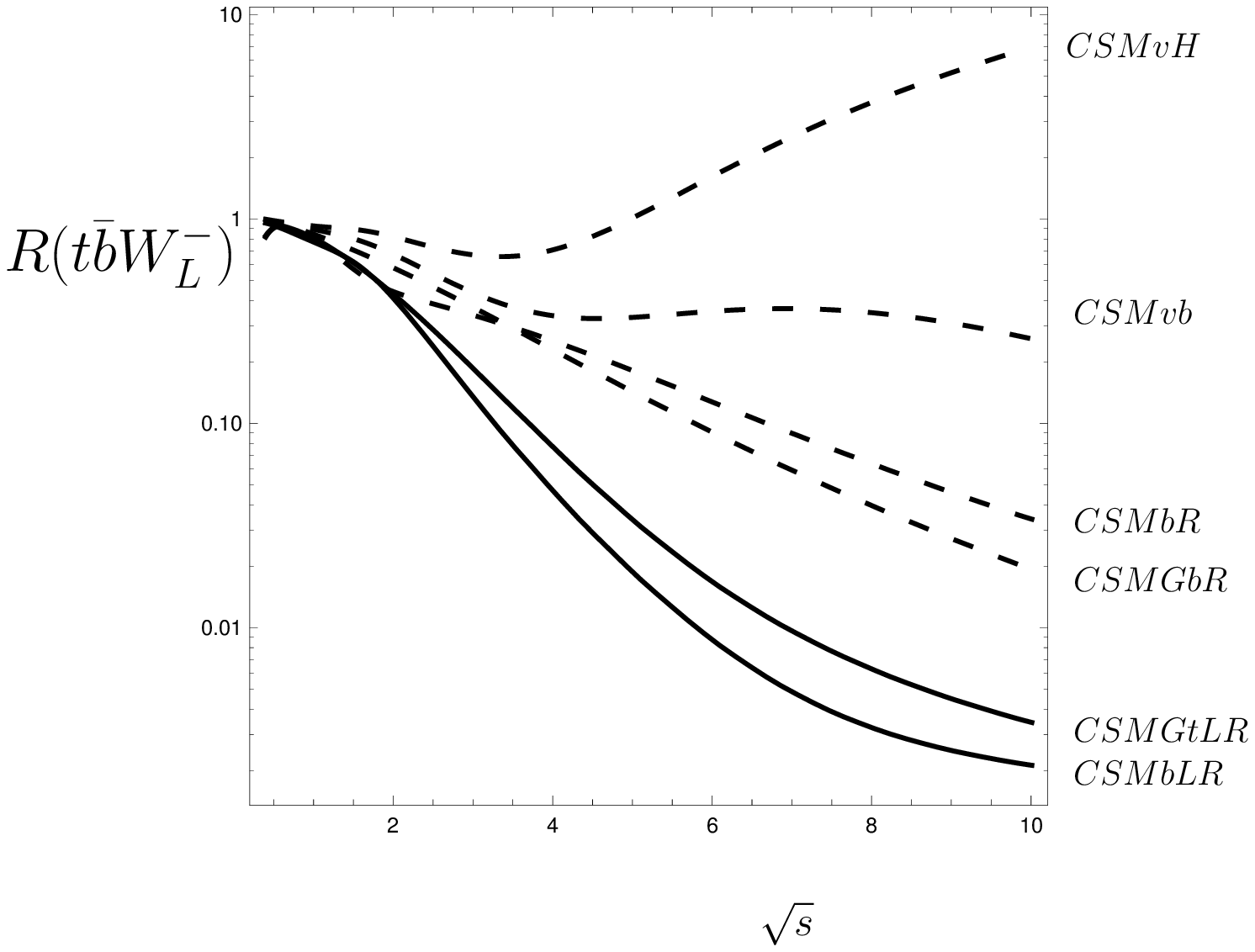, height=6.cm}
\]\\

\[
\epsfig{file=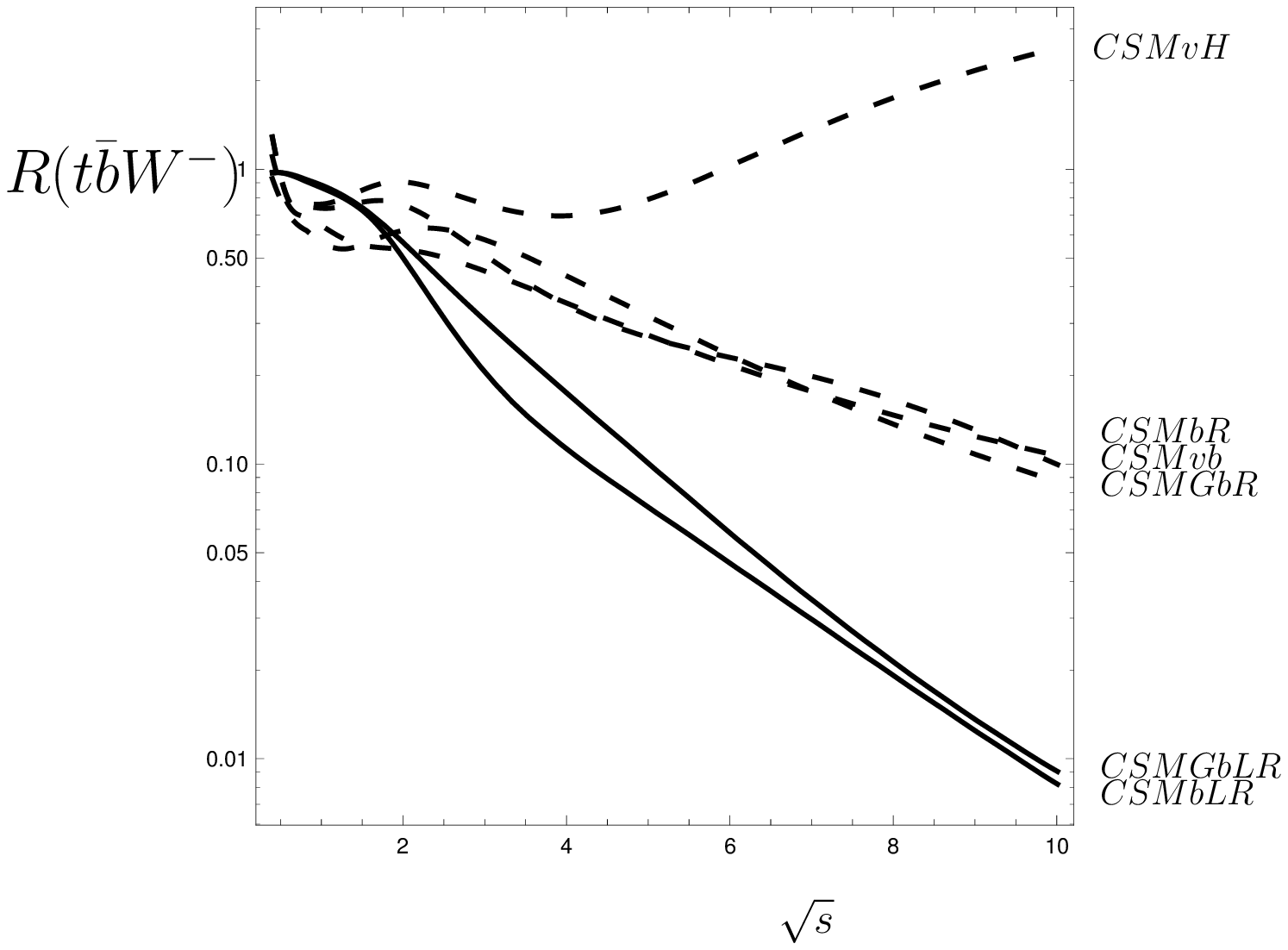, height=6.cm}
\epsfig{file=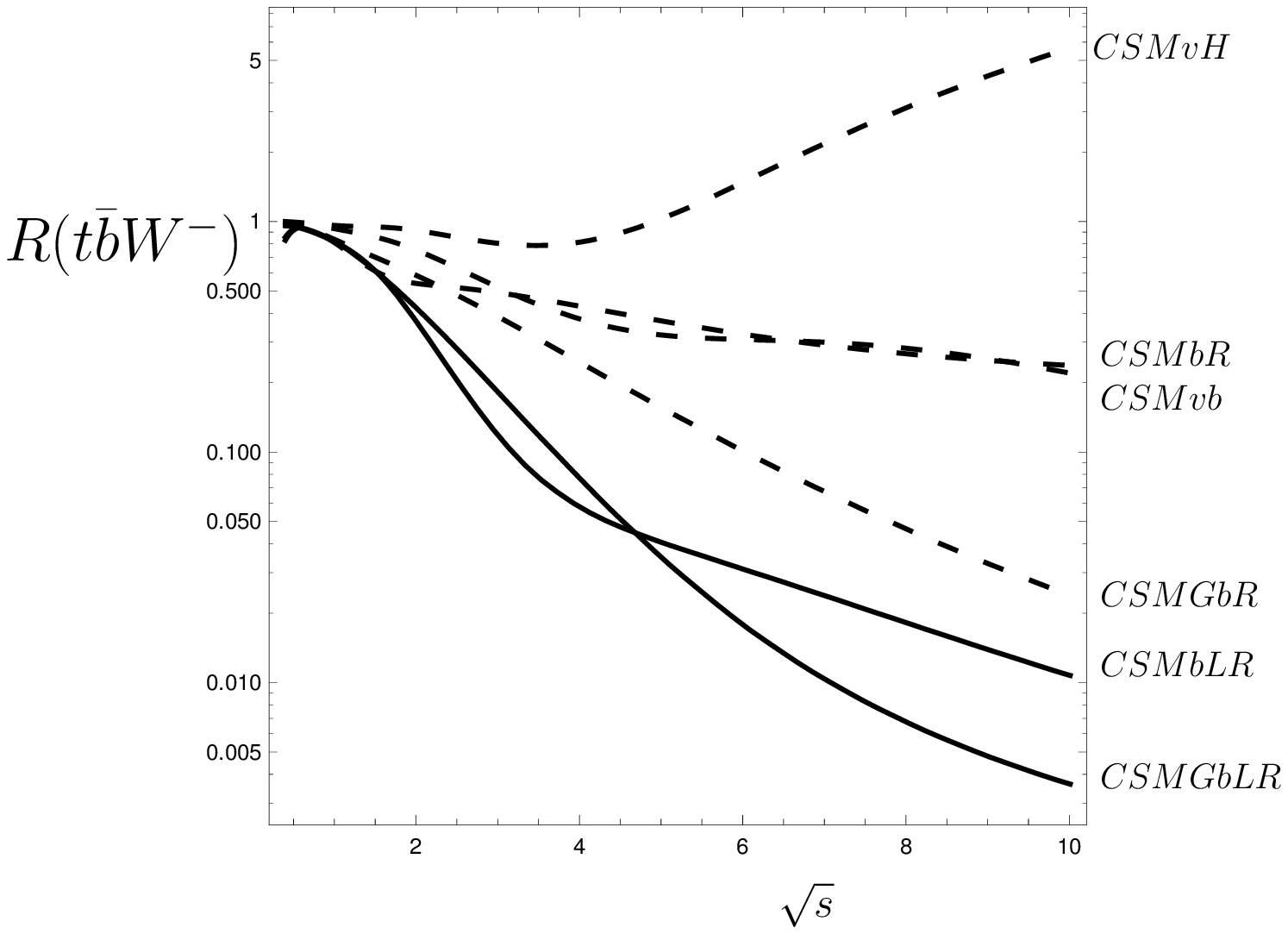, height=6.cm}
\]\\
\caption[1]  {Same ratios for $\gamma\gamma\to t\bar b W^-_L$ 
and $\gamma\gamma\to t\bar b W^-$   for elementary $b$ (left),
for full composite $b$ (right).}
\end{figure}

\clearpage

\begin{figure}[p]
\[
\epsfig{file=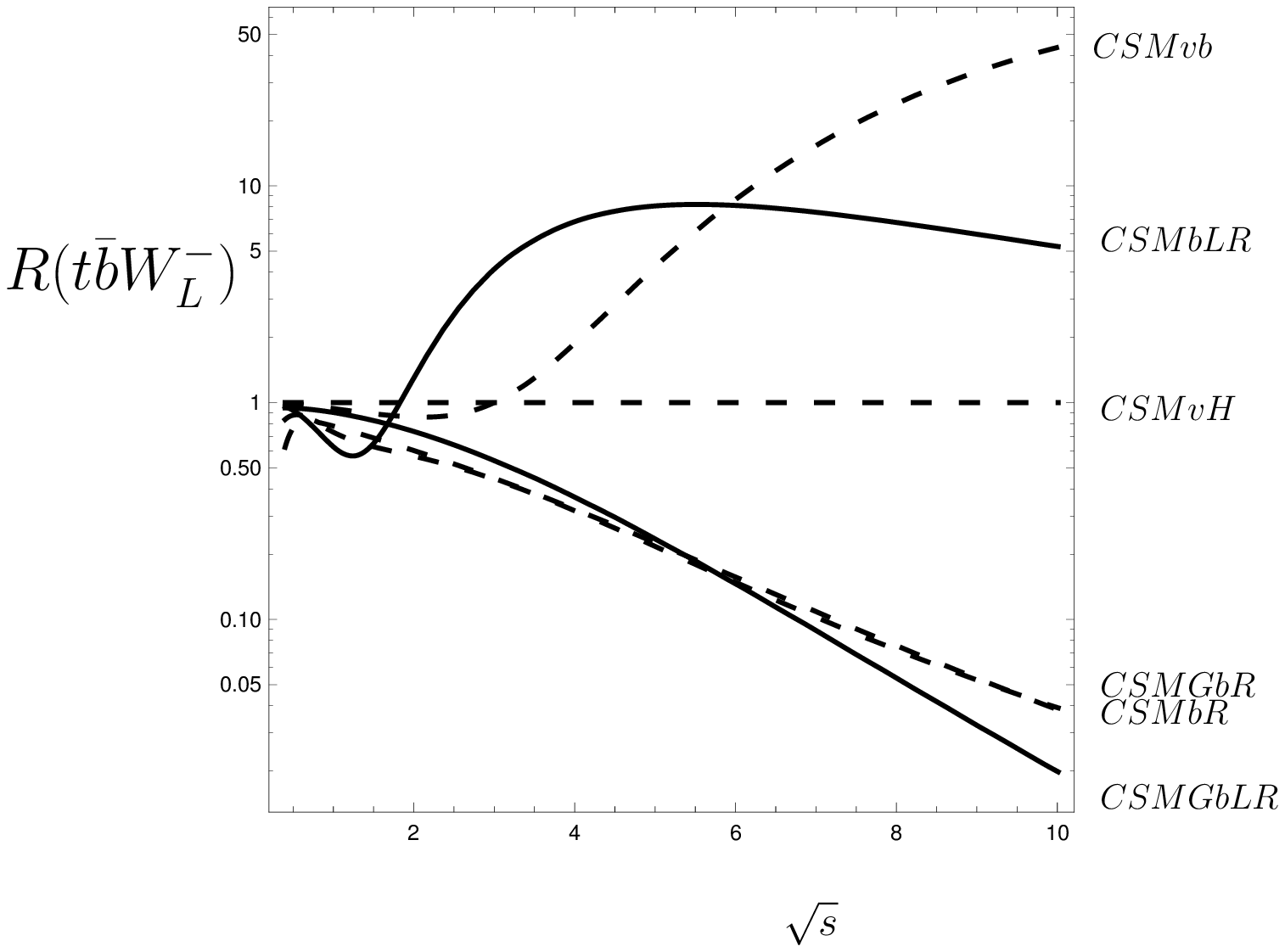, height=6.cm}
\epsfig{file=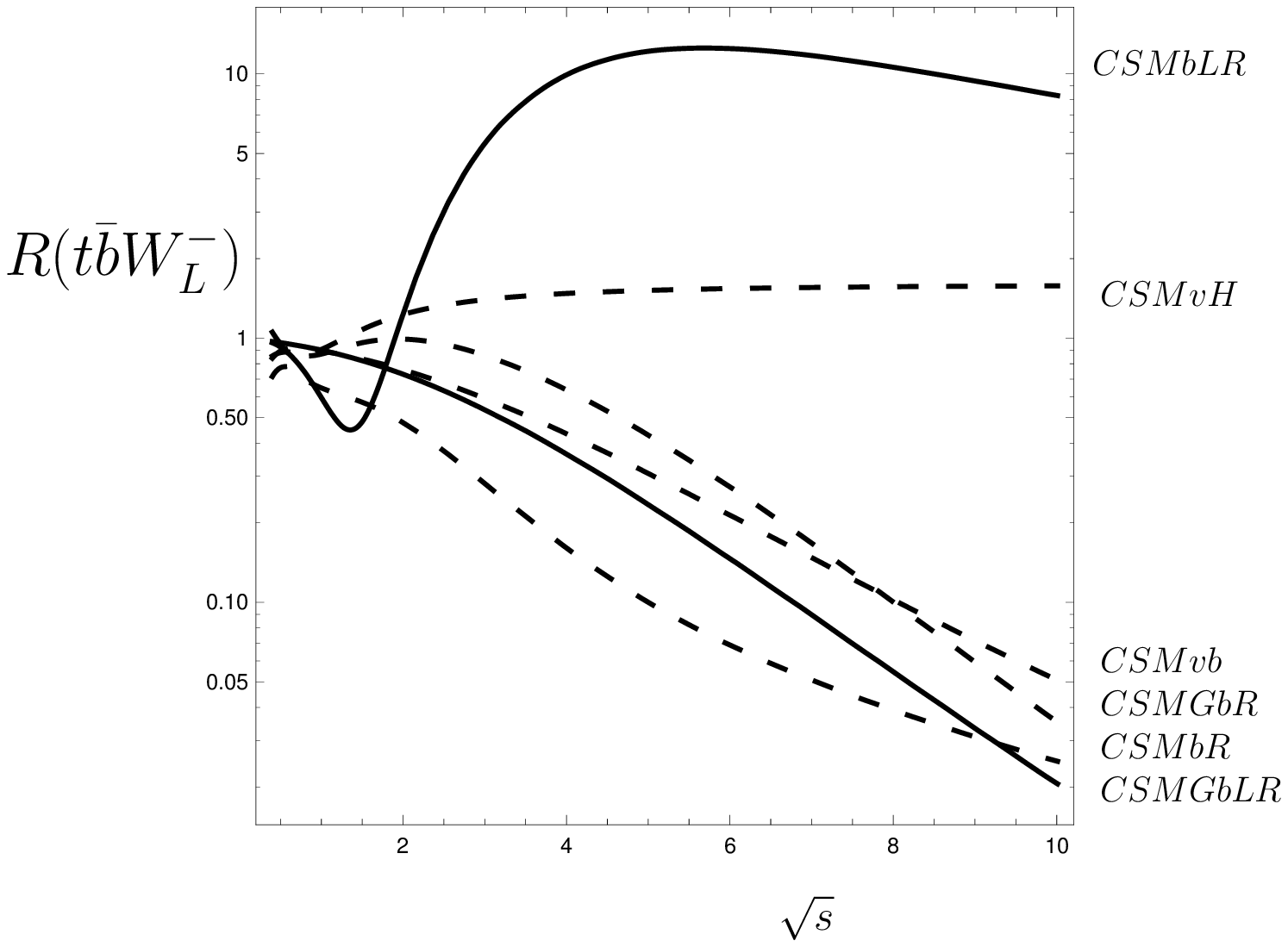, height=6.cm}
\]\\

\[
\epsfig{file=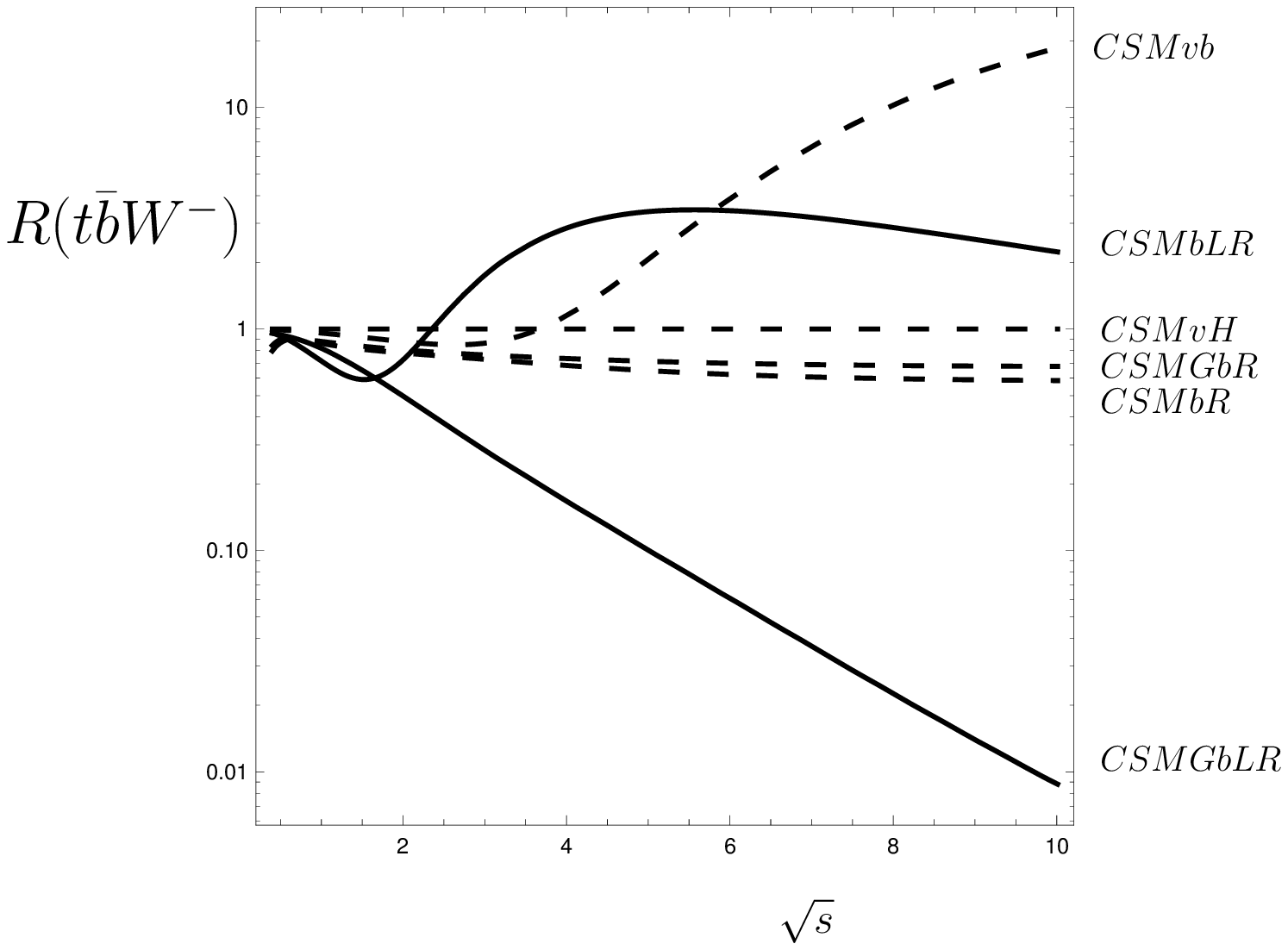, height=6.cm}
\epsfig{file=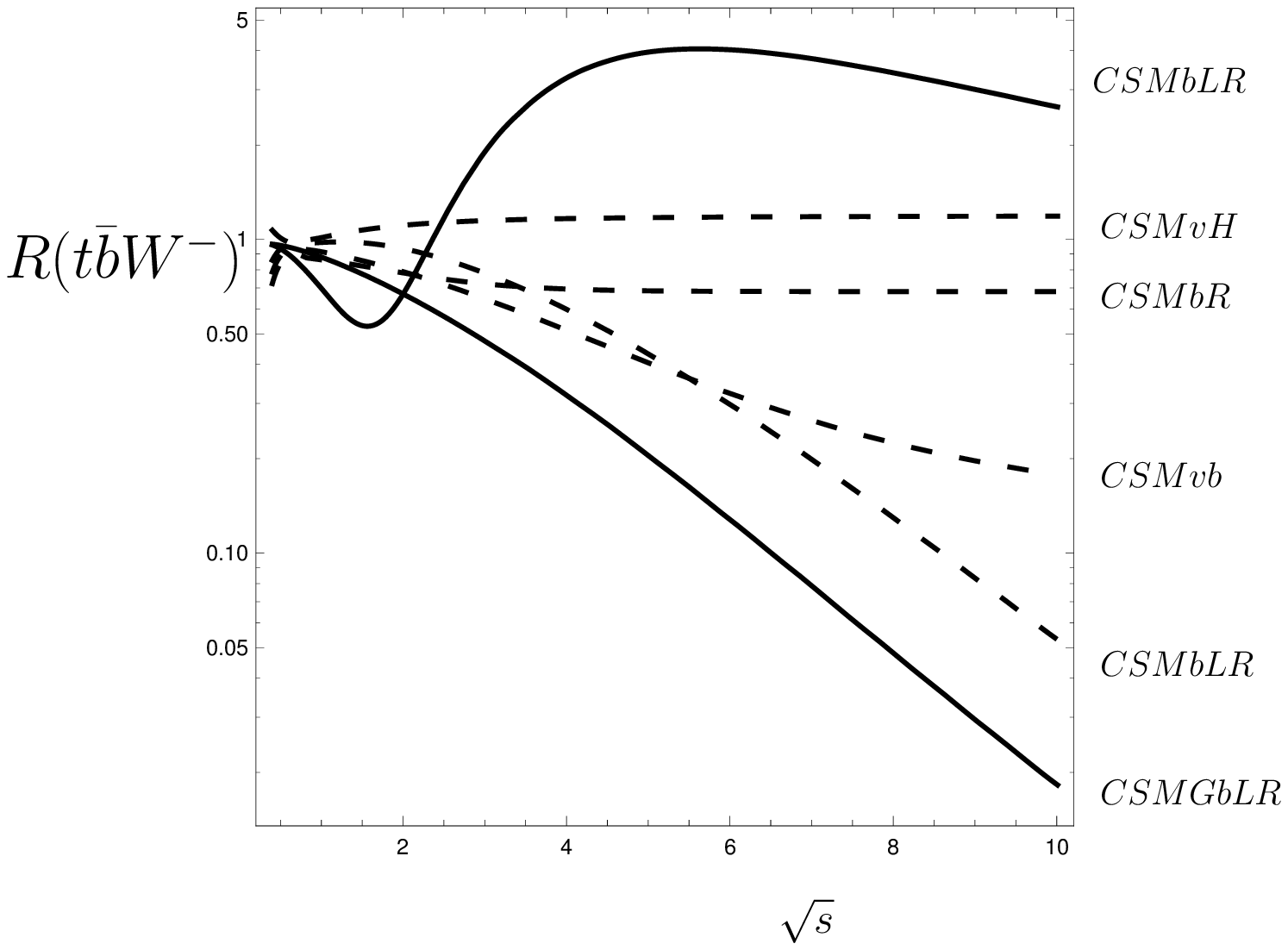, height=6.cm}
\]\\
\caption[1] {Same ratios for $gg\to t\bar b W^-_L$ 
and $gg\to t\bar b W^-$   for elementary $b$ (left),
for full composite $b$ (right).}
\end{figure}

\end{document}